\documentclass[useAMS,usegraphicx,usenatbib]{mn2e}

\newif\ifAMStwofonts
\AMStwofontstrue

\def\Vs{{V_{_{\rm seed}}}} 
\def\Vsi{{V_{_{{\rm seed},i }   }}} 
 
\def\Asj{{a^j_{_{\rm seed}}}}

\def\Psii{{P_{_{{\rm seed},i} }}} 
\def\sigint{\sigma_{\eta,{\rm int}}}
\def\ssigint{\sigma^2_{\eta,{\rm int}}}

\def\vnabla{\pmb{\nabla}}
\def\vv{\bf v}

\def\gsim{~\rlap{$>$}{\lower 1.0ex\hbox{$\sim$}}}

\def\simpropto{\lower.2ex\hbox{$\; \buildrel \propto \over \sim \;$}}
\def\ltsim{\lower.5ex\hbox{$\; \buildrel < \over \sim \;$}}
\def\gtsim{\lower.5ex\hbox{$\; \buildrel > \over \sim \;$}}
\def\ltsim{\lower.5ex\hbox{$\; \buildrel < \over \sim \;$}}
\def\gtsim{\lower.5ex\hbox{$\; \buildrel > \over \sim \;$}}

\def\pa{\partial}
 
\def\vg{v_{_{\rm g}}}
\def\ag{a_{_{\rm g}}}
\def\aitf{a_{_{\rm itf}}}

\def\vnabla{{\bf \nabla}}

\def\kms{\mbox{km\,s$^{-1}$}}

\def\dd{\,{\rm d}}

\def\nb{{n_{_{\rm b}}}}

\def\kms{\ {\rm km\,s^{-1}}}
\def\hmpc{\ {\rm h^{-1}Mpc}}

\def\dd{{\rm d}}
\def\nb{{\bar n}}
\def\ln{{\rm ln}}
\def\pa{\partial}

\def\pmb#1{\setbox0=\hbox{#1}%
\kern-.025em\copy0\kern-\wd0
\kern.05em\copy0\kern-\wd0
\kern-.025em\raise.0433em\box0}

\def\vv{\pmb{$v$}}
\def\vs{\pmb{$s$}}

\def\iras {{\it IRAS~}}

\def\Ft {{F}}

\def\jp {j^\prime}
\def\chitf{\chi^2_{_{\rm ITF} } }

\def\etal{{\it et al.\ }}

\def\simlt{\lower.5ex\hbox{$\; \buildrel < \over \sim \;$}}
\def\simgt{\lower.5ex\hbox{$\; \buildrel > \over \sim \;$}}

\def\vnabla{\pmb{$\nabla$}}
\def\vitf{v_{_{\rm itf} }}

\newcommand{\beq}{\begin{equation}}
\newcommand{\eeq}{\end{equation}}
\def\beqa{\begin{eqnarray}}
\def\eeqa{\end{eqnarray}}
\def\fixit#1{}
\def\hmpc{h^{-1}\,{\rm Mpc}}

\def\dd{{\rm d}}

\title[Gravity  versus Velocity]{Local Gravity versus Local Velocity: Solutions for $\beta$ and nonlinear bias}
\author[Davis \etal]{Marc Davis$^1$\thanks{E-mail: mdavis@berkeley.edu}, 
Adi Nusser$^2$, 
Karen L. Masters$^4$,   
Christopher Springob$^5$\newauthor
John P. Huchra$^6$, 
Gerard Lemson$^7$ \\
$^1$Departments of Astronomy \& Physics, University of California, Berkeley, CA. 94720\\
$^2$ Physics Department and the Asher Space Science Institute-Technion, Haifa 32000, Israel\\
$^3$ Smithsonian Astrophysical Observatory, Cambridge, MA 02138\\  
$^4$ Institute for Cosmology and Gravitation, University of Portsmouth,
Dennis Sciama Building, Burnaby Road, Portsmouth PO1 3FX, UK \\
$^5$ Anglo-Australian Observatory, P.O. Box 296, Epping, NSW 1710, Australia \\ 
$^6$ Harvard-Smithsonian Center for Astrophysics, 60 Garden Street,
Cambridge, MA 02138, USA\\
$^7$ Max-Planck Institute of Astrophysics, Garching DE       }         

\begin{document}


\pagerange{ \pubyear{2010}}
\maketitle
\label{firstpage}
\begin{abstract}
We perform a reconstruction of the cosmological large scale flows in the nearby Universe using two
complementary observational sets. The first, the SFI++ sample of Tully-Fisher (TF) measurements of
galaxies, provides a direct probe of the flows. The second, the whole sky distribution of galaxies in the  2MASS redshift survey (2MRS), yields
a prediction of the flows given the cosmological density parameter, $\Omega$, and   a biasing relation between mass and galaxies.
We aim at an unbiased comparison between the peculiar velocity fields extracted from the two data sets
and its implication on the cosmological parameters and the biasing relation.
  We expand the fields in a set of orthonormal basis functions, each
 representing a  plausible  realization of a cosmological velocity field
smoothed in such a way as to give a nearly constant error on the derived SFI++ velocities.
 The statistical analysis is done on the coefficients of the modal expansion of the fields by means of the basis functions.
 Our analysis completely avoids  the strong error covariance  in the  smoothed TF velocities by the
 use of orthonormal basis functions and  employs  elaborate mock data sets to extensively calibrate   the errors  in 2MRS predicted velocities.
We relate the 2MRS galaxy distribution to the mass density field by  a linear bias factor, $b$,   and include a
 luminosity dependent, $\propto L^\alpha$,  galaxy weighting.   We assess the agreement between the fields as a function
of $\alpha$ and $\beta=f(\Omega)/b$, where $f$ is the  growth factor of linear perturbations.
 The agreement  is excellent with a reasonable  $\chi^2$ per degree of freedom.
 For $\alpha=0$ , we derive  $0.28<\beta<0.37$  and
 $0.24<\beta<0.43$, respectively, at the 68.3\% and 95.4\% confidence levels (CLs).
 For $\beta=0.33$,  we get $\alpha<0.25$  and
 $\alpha<0.5$, respectively, at the 68.3\% and 95.4\% CLs.
 We set a constraint on the fluctuation normalization, finding $\sigma_8 = 0.66 \pm 0.10$, which is only $1.5\sigma$ deviant from  WMAP results.  It is remarkable that $\sigma_8$ determined from this local cosmological test is close to the value derived from the CMB, an indication of the precision of the standard model.
 

%

\end{abstract}

\pagebreak

\section{Introduction}

%
 
For 15 years, the problem of large-scale flows of galaxies  has  seen little 
 attention relative to other probes of the large scale structure in the Universe.  The data on peculiar velocities has been difficult to obtain, and the results had contradictory conclusions \citep{sw95, z02}.   They are limited to small redshifts ($\sim 100 h^{-1} \rm Mpc$)  at which distance indicators  can reliably be used.   These earlier forays into the subject led to
disagreements that few people wanted to sift through.  But in the interval,
the data has improved dramatically, thus stirring recent activity in the subject.   

Peculiar velocities are unique in that they
provide explicit information on the  three dimensional mass
distribution, and measure mass on scales of $20-50 h^{-1}$ Mpc, a scale 
untouched by alternative methods.  
Local  peculiar velocity data  is, in principle,  affluent in 
cosmological information.
Power spectra  and correlation functions 
could be derived from the data by direct calculation or  by maximum likelihood techniques \citep[e.g.][]{gd89, jk95, freu99,jusf00,brid01,abate09}.  Direct calculation of low order moments of the flow, such as the bulk
motion and the shear could also analyzed within the framework of cosmological models 
\citep[e.g.][]{feldwh10}.  Some authors claim the bulk flow of mixed catalogs of galaxies argues there are problems with 
$\Lambda CDM$ \citep{watkins09}, but recent results \citep{nuss11} show that the SFI++ catalog by itself has a large scale bulk flow that is consistent with  $\Lambda CDM$, and this analysis has smaller error bars. 
Other, perhaps more ambitious,  applications could involve an assessment of the 
statistical nature of the initial cosmological large scale fluctuations, i.e. whether gaussian or otherwise 
\citep{nd93,  berj95}.  All these analyses could be performed with peculiar velocity measurements alone. 

Here we will be concerned with a comparison of the observed peculiar velocities on the one  hand and the velocities derived from  the fluctuations in the galaxy distribution 
on the other. The basic physical principle behind this comparison is simple.
The large scale flows are almost certainly the result of the process
of gravitational instability   with overdense regions attracting material, and 
underdense  regions repelling material.  Initial conditions in the early universe
might have been somewhat chaotic, so that the original peculiar
velocity field (i.e.  deviations from Hubble flow) was uncorrelated
with the mass distribution, or even contained vorticity.  But those
components of the velocity field which are not coherent with the
density fluctuations will adiabatically decay as the Universe expands,
and so at late times one expects the velocity field to be aligned with
the gravity field, at least in the limit of small amplitude
fluctuations \citep{Peeb80, n91}.
In the linear regime, this relation implies a simple proportionality between 
the gravity field {\bf g} and the velocity field ${\bf v}_{\rm g}$,
namely   ${\bf v}_{\rm g} \propto {\bf g}~ t$ where 
the only possible time $t$ is the Hubble time.  The exact expression
depends on the mean cosmological density 
parameter $\Omega$  and is given by \citep{Peeb80},
\begin{equation}
\label{eq:vg}
{\bf v}_{\rm g}(r) = {\frac{2 f(\Omega)}{ 3 H_0 \Omega}} {\bf g}({\bf r}) \ . 
\end{equation} 
  Given complete knowledge of the mass fluctuation field $\delta_\rho({\bf r})$
over all space,  the gravity field ${\bf g(r) } $ is
\begin{equation}
\label{eq:grho}
 {\bf g}({\bf r}) =  G\bar{\rho} \int d^3 {\bf r'} \delta_\rho({\bf r'})
\frac{{\bf  r' }-{\bf  r}} { |{\bf r'}-{\bf r}|^3} \ , 
\end{equation}
where $\bar\rho$ is the mean mass density of the Universe.  
If the galaxy distribution at least
approximately traces the mass on large scale, with linear bias $b$
between the galaxy fluctuations $\delta_G$ and the mass fluctuations
(i.e. $\delta_g = b \delta_\rho$),
then from (\ref{eq:vg}) and (\ref{eq:grho}) we have  
\begin{equation}
\label{eq:vrho}
{\bf v}_{\rm g}(r) = {H_0 \beta \over 4 \pi {\bar n}}
 \sum_{i} {1\over\phi(r_i)} {\bf  r_{\rm i} -r  \over | r_{\rm i} -r|^3 }
+ {{H_0 \beta }\over 3}{\bf r}\ , 
\end{equation}
where $\bar n$ is the true mean galaxy density in the sample, $\beta \equiv f(\Omega)/b$ with $f\approx \Omega^{0.55}$ 
the linear growth factor \citep{lind05},  and where we have replaced the integral over space
with a sum over the galaxies in a catalog, with radial selection
function $\phi(r)$\footnote{$\phi(r)$ is
defined as the fraction of the luminosity distribution function observable
at distance $r$ for a given flux limit; see \citep[e.g.][]{y91}.}.
 The second term is 
for the uniform component of the galaxy distribution and  would exactly
cancel the first term in the absence of clustering within the survey volume.  Note that the result is insensitive to the value of $H_0$, as
the right hand side has units of velocity.  We shall henceforth quote
all distances in units of $\kms$. 
The sum in equation  (\ref{eq:vrho})  is to be computed in real space, whereas the galaxy catalog exists in redshift space. As we shall see in \S\ref{sec:zdist}, the modified equation,  which 
includes redshift distortions, maintains a dependence on $\Omega$ and $b$ through the 
parameter $\beta$.
Therefore,   a    comparison of the measured velocities of galaxies  to the
predicted velocities, ${\bf v}_{\rm g}(r)$, gives us measure of $\beta$.
Further, a detailed comparison of the flow patterns addresses 
fundamental questions regarding the way galaxies trace mass on large scales and 
the validity of gravitational instability theory. 


In this paper we shall make this comparison
 using the best presently available data for both the
velocity and gravity fields.
The direct comparison of the peculiar velocities  is fraught with
difficulty. Distances to individual galaxies are typically uncertain at
the 20\% level and are furthermore subject to considerable
Malmquist bias.  We shall elaborate  a method that was first presented 15 years ago \citep{DNW96} (hereafter DNW96) and which was designed to alleviate 
most of the observational biases.  But the peculiar velocity data at that time was poor and our results as well 
as those of others  \citep[e.g.][]{h95,k91,y88,sd88,branf91,ndac01,zb02} were all meant to be preliminary and none of their conclusions were  compelling.

Recently  ideal data sets have been assembled, thus allowing a new, definitive analysis of large scale flows.   The new gravity field is very well described by the nearly whole sky Two Mass Redshift Survey (2MRS)   (Huchra et al. 2005b), and the new peculiar velocity catalog is the SFI++ sample \citep{spring07, spring09}.
The 2MRS has previously been used to address the gravity field in considerable detail \citep[e.g.][]{e06,h08}, and some effort has gone into 
the comparison of the 2MRS predicted velocities versus the SFI++ measured velocities, in particular by \cite{p05, l10}.  
To date nobody has
included a proper treatment of the correlated noise in the analysis. 
Here we  shall compare the observed versus predicted  radial
velocities, taking into account a full error analysis based on a suite of elaborate mock catalogs designed to 
match the 2MRS and SFI++ data sets. 
 We shall use a refinement of the  method of
orthogonal mode expansion by \cite{ND94}
(hereafter ND94) and  \cite{ND95} (hereafter
ND95). Analysis of peculiar velocity data is
inevitably plagued  by   systematics, random measurement errors and  sparseness of the data.
The methods employed here are specifically designed to minimize these biasing, thus achieving a 
robust unbiased comparison between the measured SFI++ and the predicted 2MRS  velocities

 In \S\ref{sec:data}  we introduce the 2MRS and the SFI++ data sets and various trims that we do to ensure
 unambiguous reliable results. 
 In \S\ref{sec:vrecon} we   describe the method for extracting large scale peculiar motions from both data sets.
 We discuss the linear equation for predicting the peculiar  velocity field associated with a distribution of galaxies in 
 redshift space and review our old method 
 deriving estimates of galaxy peculiar velocities from the inverse Tully-Fisher (ITF) relation by means 
 of an expansion over orthonormal modes (basis functions). We  focus on the new refinements designed to 
 optimize the extraction of the signal from the data. 
 As has been the case in the past, mock catalogs
constructed from $N$-body simulations are essential for debugging and
calibrating the methods.  This is especially so for our application,
since the entire analysis is performed in essentially pure redshift
space. We present details of the mocks in \S\ref{sec:mocks}. 
 In  \S\ref{sec:results}  we inspect the flow fields reconstructed from the 2MRS and SF++ data, visually and statistically, demonstrating that
 differences between them are similar to those expected in the mocks. 
 We present our constraints on the cosmological parameters in \S\ref{sec:const}. 
 In the concluding \S\ref{sec:disc} we summarize our findings, discussing their implications and 
 contrasting them with other results in the literature. For readers wanting to avoid the 'how to' details, we suggest skipping \S\ref{sec:vrecon} and \S\ref{sec:mocks} but then coming back to understand how our machinery operates.

\section{New Data for the Comparison}
\label{sec:data}
\subsection{Gravity field}

Twenty years ago the only catalog of galaxy photometry with uniform
coverage over the full sky was derived from the IRAS satellite
\citep{sd88,y88}. From the point source catalog
(galaxies were unresolved in IRAS) a flux limited sample at $60\mu m$ was
constructed and redshifts were obtained for all objects to construct
the IRAS PSCz (Point Source Catalog redshift survey, Strauss et al.
1990). Among other problems, this PSCz catalog gave little weight to
ellipticals (which are dim at $60\mu m$ as this wavelength is dominated by
dusty star formation) and suffered from severe confusion in regions of high
density. However the uniform full-sky coverage was unique in enabling
the estimation of local gravity, and furthermore our local gravity
field (in a relatively low density region of the universe) is
dominated by spiral, not elliptical, galaxies and IRAS gave a fair,
although noisy, representation of the spirals.

Much larger redshift surveys do now exist e.g. SDSS \citep{abaz} and  2dF
\citep{colless03}, but few have attempted
to be complete over the whole sky as many cosmological measurements do
not require such complete surveys and a trade off has been made
between depth and sky coverage from the available telescope time and
resources. The most recent *all sky* imaging survey was the Two Micron
All-Sky Survey (2MASS, \cite{skrut}), and the 2MASS Extended
Source Catalog (XSC, Jarrett et al. 2000) extracts from that imaging a
flux limited (to K=13.5) sample of half a million extragalactic objects. The 2MASS
Redshift Survey (2MRS, Huchra et al. 2005a) is a program to obtain
redshifts for all galaxies in the 2MASS XSC to a fixed flux limit in
the K-band. The K=11.25 magnitude limited version of 2MRS consists of
~23,000 galaxy redshifts with uniform sky coverage to within 5 degrees
of the Galactic plane except towards the Galactic centre where
stellar confusion limits the catalog to +/-10 deg (Huchra et al. 2005b). A K=11.75 mag
limited version of 2MRS is almost complete, consists of ~43,000 galaxy
redshifts, and will be made available soon (Huchra et al. in prep.). 
Since the sample is K band selected, the extinction correction is modest and it is ideal for calculating local gravity. 

In the Southern hemisphere redshifts for the 2MASS galaxies were observed as part of the 6dFGS \citep{jo05, jo09},  which used the 6dF multi-fiber spectrograph on the 1.2m UK Schmidt in Siding Spring, Australia. Their ultimate product was a map of 110,256 2MASS galaxies in the Southern sky to a magnitude limit of K=12.75 mag and to within 10 deg of the Galactic plane. This survey is far deeper than the stated goal of 2MRS, but also has a higher Galactic latitude limit. In the Northern hemisphere the 2MRS builds on a strong tradition of redshift surveys at the CfA: the CfA redshift survey and ZCAT 
\citep{dav82,lgh86}.  In the absence of a northern hemisphere equivalent to the 6dF, new redshift observations are done galaxy by galaxy using the 1.2m telescope at the Fred Lawrence Whipple Observatory on Mt. Hopkins, AZ. The average density of galaxies at the magnitudes 2MRS is observing is about 1 per degree, so without a wide-field multi-object spectrograph in the Northern hemisphere this remains the most efficient way to get new redshifts.  Lower Galactic latitude galaxies in the Southern hemisphere ($ |b|> 5$) were added to 2MRS from observations at CTIO.

The version of the 2MRS which is complete to K=11.25 (consisting of
23,200 galaxies; Huchra et al. 2005, Westover 2007) has been used in
to calculate the acceleration on the Local Group by  \citep{e06a}.
The dipole estimate seems to converge to the CMB result within $60h^{-1}$ Mpc, suggesting that the bulk of the motion of the Local Group comes from structures within that distance. They also have done a dipole analysis, weighting the sample by its luminosity, rather than the counts, and find relatively minor changes.  Density and velocity fields have been calculated by 
\citep{e06} for the K=11.25  sample.  All major local superclusters and voids are successfully identified, and backside infall onto the "Great Attractor" region (at $50h^{-1}$ Mpc) is detected.

The 2MRS catalog appears to be a fair tracer of the underlying mass distribution.
   The real-space correlation lengths, $r_0$ is best fit by a regression $r_0 = (7.5 \pm 0.5) -  (3.0 \pm 0.6) {\rm log}_{10} n$, where $n$ is the cumulative number density in $10^{-3}~ h^3 $~Mpc \citep{w09}.  In contrast, \citep{ze10} report that the R band optically selected SDSS survey gives $r_0 = (6.7 \pm 0.1) - (2.0 \pm 0.1) {\rm log}_{10} n$.   
In terms of bias estimates, \citep{w09} reports  $b/b_* = 0.73 + 0.24L/L_*$  while 
\citep{no02}  state that  $b/b_* = 0.85 + 0.15L/L_* $ for the 2DF survey. 
In other words, Westover's data show the 2MRS correlations are more dependent on luminosity than are optically selected samples.
In view of this luminosity dependent result, it makes the most sense to evaluate the gravity field in a luminosity-weighted manner; it is computed below with a variety of luminosity weightings. 
\cite{w09} has also made a mock catalog for the missing galaxies at low latitude by interpolating the galaxy density above and below them in three dimensions.  We shall use this catalog as an estimate of the local mass density.

\subsection{TF sample}
Twenty years ago, the mis-calibration of full sky Tully-Fisher data was the
problem that led to very discrepant results for the determination of
$\beta \equiv \Omega/b$,  with  $\beta=0.5\pm 0.2$  \citep{DNW96} and 
$\beta=1.0\pm 0.2$  \citep[e.g.][]{dek93}.  The mistaken TF
calibration led to a large scale flow that confused both analyses, but
in the end, it was a calibration error in the Southern sky which made
a false large-scale flow \citep{wcf97}.  In one analysis this led to a higher 
$ \chi^2 $ than was acceptable, and in the other it led to a biased result.  

For the analysis below, we use the recently completed survey of spiral
galaxies with I-band Tully-Fisher distances, SFI++  \citep{mas06, spring07, spring09}, 
which in turn builds on the original Spiral Field
I-band Survey \citep{ gio94, gio95, hg99} and
Spiral Cluster I-band Survey \citep{gio97a, gio97b}.  We use
the published SFI++ magnitudes and velocity widths, and derive our own
peculiar velocities, rather than taking the published distances as given.
We use the SFI++  catalog  as it includes several datasets to give full sky coverage.   It is not essential for our analysis that the peculiar velocity sample have uniform sky coverage, but they must have a uniform calibration.

The other major Tully-Fisher catalog was published by \citep{t08}.  This survey is restricted to $cz<3000$ km/s, and includes many of the same galaxies in that redshift range as SFI++.  Tully also make use of a different algorithm for measuring spectral line widths, which are not easily comparable to the values derived for SFI++.  So while one could in principle combine the two catalogs for this analysis, the small potential gain in sample size is not enough to justify the resulting heterogeneity in observational methods and data analysis.

We shall use the inferred distances, as well as redshifts,  to derive an estimate of peculiar velocity for each galaxy. Correlation analysis \citep{bor00} indicates that peculiar velocities in the SFI++ behave as expected for $\Lambda CDM$ models. (see esp. \citep{bran01,freu99, daCostaNuss,  feldw08})

\begin{figure} 
\centering
\includegraphics[ scale=0.4 ,angle=00]{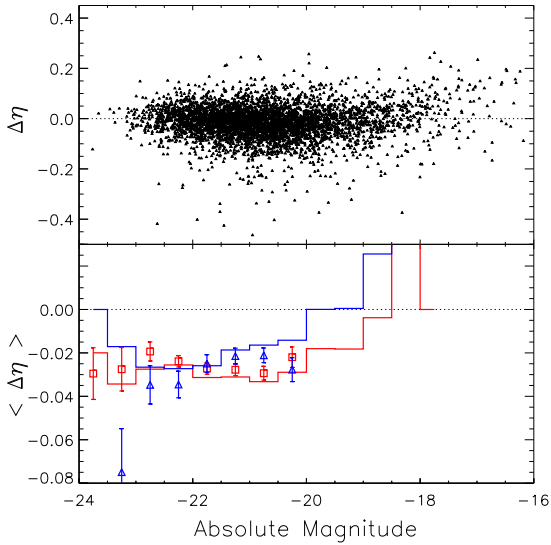}
\caption{ top: The scatter of the $\eta-M_I$ relationship, where $\gamma=0.12$  Bottom: The mean $\Delta\eta$ 
in .5 magnitude bins of  the raw distribution (recall that $\eta=$ log(W)). The red is for galaxies with $cz > 5000$ km/s, while the blue is for $cz < 5000$ km/s. Note the change where $M_I > -20$, and also how the red and blue curves appear to have  different  TF relations.  The histograms are for the raw data, with no flow model, while the points, with $1\sigma$ error bars, are the mean values of $\eta-M_I$ after the fit.  The nearby and more distant galaxies now have identical TF relations.}
\label{fig:eta_M}
\end{figure}

In the analysis below, we shall use the inverse of the Tully-Fisher (ITF)
relationship, as given in equation \ref{eq:ITF}.  We begin by drawing the published
magnitudes, velocity widths, and redshifts from \cite{spring07, spring09}.  We
include all field, group, and cluster galaxies, which leaves us with an
initial sample of 4859 galaxies.  Galaxies in groups and clusters are
treated as individual objects, though the redshifts for template cluster
galaxies are replaced by the systematic redshift of the cluster.  Following
Giovanelli et al. (1997a), we brighten the magnitudes of Sb galaxies by 0.10
magnitudes and brighten the magnitudes of spirals earlier than Sb by 0.32
magnitudes, while leaving types later than Sb unchanged.  This is done in
order to account for subtle differences in the TF relation of different
spiral subclasses.
We  select only objects with inclination $i > 45$ deg to ease
problems with inclination corrections.  The data must be transformed to the
LG frame, and galaxies with  $cz<200$ km/s  are deleted. All the analysis is done in the LG frame as the boundary conditions then simply becomes $v_g \rightarrow 0$ and $v_{itf} \rightarrow 0$.


The few galaxies with large residual $\Delta\eta$ ($\Delta\eta$ is the residual from equation \ref{eq:ITF}; see figure \ref{fig:eta_M}) are  sufficiently deviant to be a worry for statistics which depend on data with a compact core and no long tails.  The typical outlying object is not unusually nearby in redshift, and peculiar motions cannot  be the explanation.  The vast majority of SFI++ galaxies have well behaved TF relationships; perhaps the outliers are undergoing a merger?  There are large negative $\Delta\eta$ outliers, but few corresponding large positive $\Delta\eta$ outliers, and this is resolved by
clipping the outliers at $| \Delta\eta | > 0.20$.

The top panel of figure \ref{fig:eta_M} shows the distribution of $\Delta\eta$ {\it before} galaxies have had their magnitude changed because of peculiar velocities;  in  
the bottom panel is shown the result of averaging the data into .5 mag bins, where the red histogram is for galaxies at $cz > 5000$ km/s, and the blue histogram is for those with $cz < 5000$ km/s, in the raw data.  The red and blue points, plus 
$1-\sigma$ error bars, are the average $<\Delta\eta>$  versus $M$ {\it after} the best flow model, described 
in \S\ref{sec:refine}, is fit to the data. The blue point with $M<-23$ is deviant, but it only represents 11 galaxies, compared to an average of $170$ galaxies in the other bins.   Note that the zero point of the $<\Delta\eta>$ behavior makes no difference; only the slope, the constancy of $<\Delta\eta>$ versus $M$ is important. Compared to the different slopes before the flow corrections are applied, the TF relation is now identical in the foreground versus the background of the SFI++ data.
This figure is for illustration only, as the data is not binned during the fitting process. 

The bend in the TF relationship at $M=-20$ is known to be a result of the reduced mass in the Baryonic Tully-Fisher relation \citep{stark09,guro10}.  We are missing the data to straighten out the curve, and since
the ITF method is easiest to apply if there is a linear relationship between  $\eta$ and $M$, 
we simply delete all galaxies with $M > -20$ from further consideration.  After all these cuts, we are left with 2830 spiral galaxies with $200 < cz_{LG} < 10000$ km/s. 

The raw distribution of $\Delta \eta$, after limiting the sample and fitting the best linear curve, is found to approximately fit a gaussian  with 
$\sigma = 0.059$.   This is the dispersion with {\it no flow model} applied.  
The gaussian width to the distribution is $\sigma = 0.0558$ after the flow model is applied.  The small decrease is  limited by the intrinsic, dominant noise of the TF relation. This noise has numerous causes, such as the uncertainty in the inclination correction of the SFI++ galaxies, or small variations in the outer limits of the rotation curves of the galaxies.

 The following sections,  \S\ref{sec:vrecon} and \S\ref{sec:results}, explain the machinery for effecting this reduction.


\section{Reconstruction of Peculiar Velocities}

\label{sec:vrecon}

In this section we outline our method described in ND94, ND95 and DNW96  for deriving the smooth peculiar velocities of galaxies from
an observed distribution of galaxies in redshift space and, independently,
from a sample of spiral galaxies with measured circular velocities 
$\eta$ and apparent magnitudes $m$.

\subsection{Peculiar Velocities from the Distribution of Galaxies in Redshift Space}
\label{sec:zdist}
There are several   methods
for generating peculiar velocities from redshift surveys, 
using linear  \citep[e.g.][]{Fisher95b}  and non-linear relations 
 \citep[e.g.][]{Peeb80,cc98,NussBranch,Frisch, Ensslin09}
Here we restrict ourselves to large scales where linear-theory is applicable.
We will use the  method of ND94  for reconstructing velocities from the 2MRS. This method is 
is particularly convenient, as it is easy to implement, fast,
and requires no iterations. Most importantly, this redshift space
analysis closely parallels the ITF estimate described below. 
We next present  a very brief summary of the
methodology.

We follow the notation of DNW96. The comoving redshift space coordinate and 
the comoving peculiar velocity relative to the Local Group (LG) are, respectively, denoted by 
 $\vs $ (i.e. $s= cz/H_0$) and $\vv(\vs)$.
To first order, the peculiar velocity is irrotational
in redshift space \citep{chodnuss} and can be expressed  as $\vv_g(\vs)=-\vnabla\Phi(\vs)$ where  $\Phi(\vs)$ is a potential function.
As an estimate of the fluctuations in the fractional density field $\delta_0(\vs)$ traced by the discrete distribution 
of galaxies in redshift space  we consider,
\begin{equation}
\delta_0(\vs)= 
{1\over {(2\pi)^{3/2}\nb \sigma^3}}\sum_i {{w(L_{0i})}\over {\phi(s_i)}} 
\exp\left[-{{\left(\vs - \vs_i\right)^2}\over
{2\sigma^2}}\right] -1 \quad . 
\label{eq:deldef}
\end{equation}
where $\bar n=\sum_i w(L_{0i})/\phi(s_i)$ and $ w$ weighs  each   galaxy according to its estimated luminosity, $L_{0i}$. 
The 2MRS density field is here smoothed by a gaussian window 
with a redshift independent width, 
$\sigma=350\kms$. This is in contrast to DNW96 where the \iras 
density was smoothed with a width proportional to the mean particle separation.
The reason for adopting a constant smoothing for 2MRS is its dense sampling which is nearly four time 
higher than  \iras.
We emphasize that the coordinates ${\bf s}$ are in {\it
observed redshift} space, expanded in a galactic reference frame.  The only
corrections from pure redshift space coordinates is the
collapse of the fingers
of god of the known rich clusters prior to the redshift space smoothing
(Yahil \etal 1991).
Weighting  the galaxies in equation (\ref{eq:deldef})  by
the selection function and luminosities evaluated at their redshifts rather than the
actual (unknown) distances yields a biased estimate for
the density field.  This bias   gives rise to Kaiser's rocket effect \citep{kais87}.  

To construct the density field,  equation \ref{eq:deldef}, we volume limit the 2MRS sample to 3000 km/s, so that $\phi{(s<3000)}= 1$, resulting in $\phi{(s=10000)}=0.27$ \citep{w09}.  In practice, this means we delete galaxies from the 2MRS sample fainter than $M_* + 2$.  Galaxies at 10,000 km/s therefore have $1/\phi = 3.7$ times the weight of foreground galaxies in the generation of the velocity field, $v_g$.     

If we expand the angular dependence of $\Phi$ and $\delta_0(\vs)$
redshift space   in 
spherical harmonics 
in the form, 
\begin{equation}
\Phi(\vs)=
\sum_{l=0}^{\infty}\sum_{m=-l}^l \Phi_{lm}(s)Y_{lm}(\theta,\varphi)
\end{equation}
and similarly for $\delta_0$, then, to first order,
$\Phi_{lm}$ and $\delta_{0lm}$ satisfy,
\begin{eqnarray}
\label{eq:phis}
{1\over {s^2}} {\dd \over {\dd s}}
\left(s^2 {{\dd \Phi_{lm}} \over {\dd s}}\right)
&-&{1 \over {1+\beta}}{{l(l+1) \Phi_{lm}} \over {s^2}}\\
&=&{\beta \over {1+\beta}} \left(\delta_{0lm} - 
{ \kappa(s) } { {\dd \Phi_{lm}} \over {\dd s}}\right)\; , \nonumber
\end{eqnarray}
where 
\begin{equation}
\kappa=\frac {\dd \ln\phi}  {\dd  s} - \frac{2}{s} \frac{d \ln w(L_{0i})}{d \ln L_{0i}} 
\end{equation}
represents the  correction for the bias introduced by 
 the generalized Kaiser rocket effect.  As emphasized by
ND94, the solutions  to equation (\ref{eq:phis}) for the monopole ($l=0$) and the dipole
($l=1$) components of the radial peculiar velocity in the LG frame 
are uniquely determined by specifying vanishing velocity at the origin.  
That is, the radial velocity field at redshift $\bf s$,  when expanded to
harmonic $l \le 1$, is not influenced by material at redshifts greater
than $\bf s$.  

In this paper, we shall consider solutions as a function 
of $\beta$ and the parameter $\alpha$ defining a power law form $w_i\propto L_i^\alpha$ for
the galaxy weights.

\subsection{Peculiar Velocities from the Inverse Tully-Fisher relation}
\label{sec:itfd}
Given a sample of galaxies with measured circular velocity
parameters, $\eta_i  \equiv {\rm log}\omega_i$, linewidth $\omega_i$, apparent magnitudes $m_i$, 
and redshifts $z_i$, the goal is to derive an estimate for the smooth
underlying peculiar velocity field.  We assume that the circular
velocity parameter, $\eta$, of a galaxy is, up to a random
scatter, related to its absolute magnitude, $M$, by means of a
linear {\it  inverse} Tully-Fisher (ITF) relation, i.e.,
\begin{equation}
\label{eq:ITF}
\eta=\gamma M + \eta_0 .  
\end{equation}

One of the main advantages of inverse TF methods is that  samples selected by magnitude, as most are, will be minimally plagued by   Malmquist bias effects when analyzed in the inverse
direction  \citep{schechter, a82}.
We write the absolute magnitude of a galaxy, 
\begin{equation}
M_i = M_{0i} + P_i
\end{equation}
where
\begin{equation}
M_{0i} = m_i + 5{\rm log}(z_i)-15 
\end{equation}
and 
\begin{equation}
P_i = 5{\rm log}(1- u_i/z_i)
\label{eq:P_i}
\end{equation} 
where $m_i$ is the
apparent magnitude of the galaxy, $z_i$ is its redshift 
in units of $\kms,$
and $u_i$ its radial peculiar velocity in the LG frame.

 ND95 base a velocity  model on spherical harmonics and  spherical Bessel functions,  for galaxies distributed over the sky to 6000 km/s. With the 2MRS we extend the gravity field to 10,000 km/s.
In general, one can write the
function $P_i$ in terms of an expansion over $j_{_{\rm m}}$ orthogonal basis functions, $\Ft^j_i$,
\begin{equation}
\label{eq:pF}
P_i =\sum_{j=0}^{j_{_{\rm m}}} a^j \Ft^j_i 
\end{equation}
with orthonormality conditions,
\begin{equation}
\label{eq:Forth}
\sum_{i=1}^{N_g} \Ft^j_i  \Ft^{\jp}_i= \delta_K^{j , {\jp}}  
\end{equation}
and the  zeroth mode defined by $\Ft^0_i=1/\sqrt N_g$, where $N_g$ is 
the number of galaxies in the sample. The  mode $\Ft^0$
describes a Hubble-like flow in the space of the data set which is 
degenerate with the zero point of the ITF relation.
Here we 
 set $a^0=0$, which removes the Hubble-like flow from the gravity field, below. 
The best fit mode coefficients, $a^j$, the slope, $\gamma$, and  the zero point
$\eta_0$, are found by  minimizing the $\chi^2$ statistic
\begin{equation}
\label{eq:chi}
\chitf =\sum_{i=1}^{N_{\rm g}}
  \frac{\left(\gamma M_{0i}+\gamma P_i+\eta_0-\eta_i\right)^2}{\ssigint}\; ,
\end{equation}
where $\sigint$ is the $rms$ of the intrinsic scatter in $\eta$
about the ITF relation, and $N_{\rm g}$ is the number of galaxies in the sample.
Given the orthonormality condition, the solution to the equations $\partial \chitf/\pa a^j=0$, 
 $\partial \chitf/\pa \gamma=0$ and  $\partial \chitf/\pa \eta_0=0$ is straightforward.
 Thanks to the orthonormality condition,  the covariance matrix $<\delta a_j \delta a_j'>$ of the errors in $a^j$ 
  is diagonal with   
 \begin{equation}
 \label{eq:siga}
 \sigma_a=<(\delta a^j)^2>^{1/2}=\frac{\sigint}{\gamma} \; .
 \end{equation}
 This lack of covariance of the errors in the coefficients is most rewarding as it makes the 
 ITF error analysis exceptionally simple.
 Therefore,  statistical assessment of the match between the data will be done at the level of the modes
 rather than the peculiar velocities.
 The interested reader will find details in ND95 and DNW96.

 \subsection{The orthonormal basis functions} 
The choice of radial basis functions for
the expansion of the modes can be made with considerable latitude.  The 
functions should obviously be linearly independent, and close to orthogonal
when integrated over volume. They should be smooth and close to a complete
set of functions up to a given resolution limit.   ND95  chose spherical
harmonics $Y_l^m$ for the angular wavefunctions and the 
derivatives of spherical Bessel functions for the radial basis functions, 
motivated by the desire to use functions which automatically satisfy
potential theory boundary conditions at the origin and the outer boundary.
That is, they chose 
\begin{equation}
P(y,\theta,\phi)=\sum_{n=0}^{n_{max}}\sum_{l=0}^{l_{max}}\sum_{m=-l}^{m=l}
{a_{nlm} \over y}\left(j^{\prime}_{l}\left(k_ny\right)-c_{_{l1}}\right)
Y_{lm}\left(\theta,\phi\right) \; .
\end{equation}

The function  $y(z)$ is designed to compress the distance scale, increasing the smoothing scale of the mode to deal with increased noise at large distances. For this analysis we use
\begin{equation}
\label{eq:yz}
y(z) =  \sqrt{{\rm log}(1+(z/z_s)^2)}
\end{equation}
where $z_s = 5000$ km/s.
The constant $c_{l1}$ is non-zero for the dipole term only and ensures that
$P=0$ at the origin, and is non-zero at the outer boundary.
Details of how the orthogonalized functions $\tilde F_i^j$ are derived
from this expansion are given in ND95.

The spherical harmonics are expanded  to a maximum   $n=5$ and $l=3$, except we delete the $n=1$ mode for $l=0$   as this mode can be confused with the false Hubble flow described in the next section. We also include an external quadruple, distinct from the internal  quadruple, to describe the gravity induced by material at distances $cz>10000\kms$. Summing over the values  of $m$, that makes a total of 72 modes fit toward reducing the $\chi^2$  of equation \ref{eq:chi}.  The use of $y(z)$ is designed to allow the radial resolution to degrade with distance; for example,  the  $n=5$ modes have a wavelength in the radial direction of $3000\kms$ at $cz=8000\kms$ and a wavelength of $1300\kms$ at $cz=1000\kms$.

\subsection{ Expanding the 2MRS gravity field}

In order to assess the match between the velocities by
means of the expansion coefficients and to ensure that both fields 
are smoothed similarly, the 2MRS predicted velocities $\vg$ must also be described by 
an expansion over the basis functions used in the $\vitf$ model.
Using  the machinery for computing a gravity field described in \S\ref{sec:zdist}, 
one can generate a linear theory predicted peculiar velocity
$\vg$ for any point in space as a function of its redshift for
any value of $\beta$.  We must ensure that the smoothing scales
of the ITF  and 2MRS predicted peculiar velocities are matched
to the same resolution. Therefore we expand $\vg$ in terms of the
modes used in the velocity model.  Because of the
orthonormality, we can write the mode coefficients as
\begin{equation}
\label{eq:a_g}
a_{_{\rm g}}^j = 5 \sum_{i=1}^{N_{\rm g}} {\rm log}\left(1 - {(v_{_{{\rm g},i }} - H' cz_i)\over cz_i} 
\right) \Ft^j_i
~ , 
\end{equation}
where the $H'$ term is a correction for Hubble flow and the summation
index $i$
is restricted to be
over the positions of the same galaxies in the ITF expansion.

This procedure will filter out  fluctuations  that are not
described by the resolution of our basis functions. 
We do not include any mode such that $P_i = $ constant, which would be a pure Hubble flow. 
In the fitting for the ITF modes, pure Hubble flow
is absorbed into a shift of the zero point $\eta_0$ and the
orthogonality is ensured.  Within a given set of test points
occupying a volume smaller than that used to define the gravity
field, it is possible for $v_g$ to have a non-zero value of
Hubble flow $H'$, which must be removed from $v_g$ before we
tabulate the mode coefficients.  That is,  we tabulate the mean
Hubble ratio
\begin{equation}
H' =  \frac{\sum_{k=1}^{N_{\rm g}} v_{g,k}  cz_k }{ \sum_{k=1}^{N_{\rm g}}  v_{g,k} ^2}
\end{equation}
and subtract it from the predicted field $v_g$.  This ``breathing
mode" which mimics a Hubble flow is not
trivial in amplitude, and can be a 10\% correction on the effective Hubble
constant within simulated catalogs. This mode is cosmologically expected to a modest degree, but a bigger portion of the effect is caused by error in the determination of $\overline{n}$, which we estimate by assuming the weighted counts within 12,000 km/s is the mean value.

\subsection{Refinements of  the ND95 functions}
\label{sec:refine}
The \cite{ND95}  reconstruction of the  base functions can only provide a rough estimate of 
the spatial distribution of galaxies in the TF data. It does  not guarantee  that the signal to noise in the 
filtered fields is uniform all over the  sample. 
The ND95  expansion yields reliable TF velocities of   nearby galaxies,  
but very noisy estimates at larger distances. 
Moreover, it is difficult to achieve a desired  resolution  as a function of redshift and 
 to ensure equal resolutions in the radial  and  angular  directions.
  The ND95 method expands the observed velocities in terms of harmonic functions, but the 
individual harmonic modes are not  regularized and may acquire unrealistically  large amplitudes,  depending on the spatial coverage of the data. 

We aim here at generating base functions which are themselves smoothed 
with a variable isotropic smoothing window designed to yield a constant signal to noise in the 
estimated $\vitf$. 
We construct these new basis  functions with the help 
of the ND95 orthogonal functions denoted here by $F^{\rm  ND}$.
Suppose a single radial velocity field, $\Vs$,  with the appropriate variable smoothing 
has been found.  
We term $\Vs$ the {\it seed} field as the new modes will stem from it. As will be described below this field will be chosen as 
the 2MRS predicted velocity field, but any other field representing a viable velocity field could serve as $\Vs$. Given 
  \begin{equation}
  \Psii=5{\rm log}(1-\Vsi/cz_i)
  \end{equation}
   (where $i$ refers to galaxies in the  TF sample)
we expand $\Psii$ in  base functions constructed according to ND95, 
\begin{equation}
\Asj=\sum_i \Psii F^{NDj}_i\; .
\end{equation}
Here, the number of the modes $F^{ND}$ is sufficiently large so that the inverse transformation $\sum_j \Asj F^{NDj}_i$
reproduces 
 $\Psii$.  In practice we use about 1400 ND95 modes (we go to $l=17$).  

\begin{figure}
\centering
\includegraphics[ scale=0.45,angle=00]{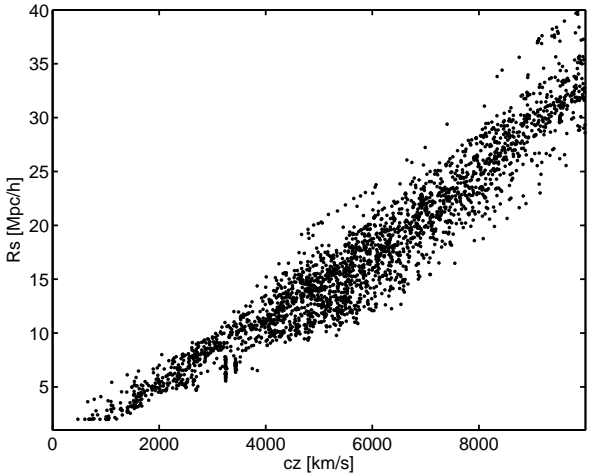}
\caption{ The  width of the gaussian  smoothing gaussian window 
versus galaxy redshifts in the SFI++ sample. The scatter  reflects the angular variations in the 
density of galaxies. }
\label{fig:Rs}
\end{figure}

We then   form additional fields,  $P_\alpha$, according to 
\begin{equation}
P_{\alpha, i}=\sum_j{\cal R}^j_{\alpha} \Asj F^{NDj}_i \; ,
\end{equation} 
where $\cal R$ is a set of normally distributed
random numbers with zero mean and standard deviation of unity. 
This reconstruction of the additional 
fields preserves the ``power" in the modal expansion and randomizes  the phases.  
So far all these fields,  $P_{\alpha,i}$,  are unfiltered and may contain non-linear small scale fluctuations.  
Therefore, we smooth all fields  $P_{\alpha,i}$ according to 
\begin{equation}
 P^{\rm smooth}_{\alpha,i}=\sum_{all \; galaxies} P_{\alpha,i'} W(s_{i,i'},R_{s,i'})\; ,
\end{equation}
where  $s_{i,i'} $ is distance (in redshift space) between the galaxies $i$ and $i'$,  and  $W$ is a gaussian window
of width $R_{s,i}$  which depends on the galaxy $i$. 
The smoothing width $R_{s,i}$ is tuned such that the expected
error in the ITF velocity of galaxy $i$ is $\sim 150\kms$ and therefore it depends on the redshift and the 
local density of galaxies near $i$. 
 The smoothing length at the positions of galaxies in the SFI++ sample is shown in figure (\ref{fig:Rs}); it  varies roughly   linearly with redshift, ranging from 
 $1\hmpc$ nearby up-to $30\hmpc$ for galaxies at redshifts $\sim 10000\kms$.  
The new basis functions  are then obtained by  orthonormalizing  $P^{\rm smooth}_{\alpha,i}$. 
 We refer to the new 
functions by the standard notation $F^j_i$, with $F^1_i$ being the smoothed $\Psii$.  These new modes will 
be used in the expansion given in equation (\ref{eq:pF}).  In  \S\ref{sec:TheITF} we describe how we 
determine the number of modes, $j_{_{\rm m}}$, to be used  in the expansion. 
The seed  field $\Vs$ could be constructed by interpolating any unfiltered cosmological velocity field
 on the positions of the 
galaxies in the SFI++  sample.
Nevertheless,  we could improve on this by constructing $\Vs$ from the 
unfiltered 2MRS velocities given directly by the solution to equation  (\ref{eq:phis}). 
 In practice, we use unfiltered $\vg$ obtained with $\beta=0.2$. The choice of $\vg$ for $\beta=0.2$ is arbitrary; the predicted field  with any other $\beta$ could be used. 
 If the $\Vs \sim \vg$ and $\vitf$  both describe  the same underlying velocity field then 
the additional modes should mainly reflect the covariance of the errors 
between the two fields.

\begin{figure*}
\centering
\includegraphics[ scale=1.1 ,angle=00]{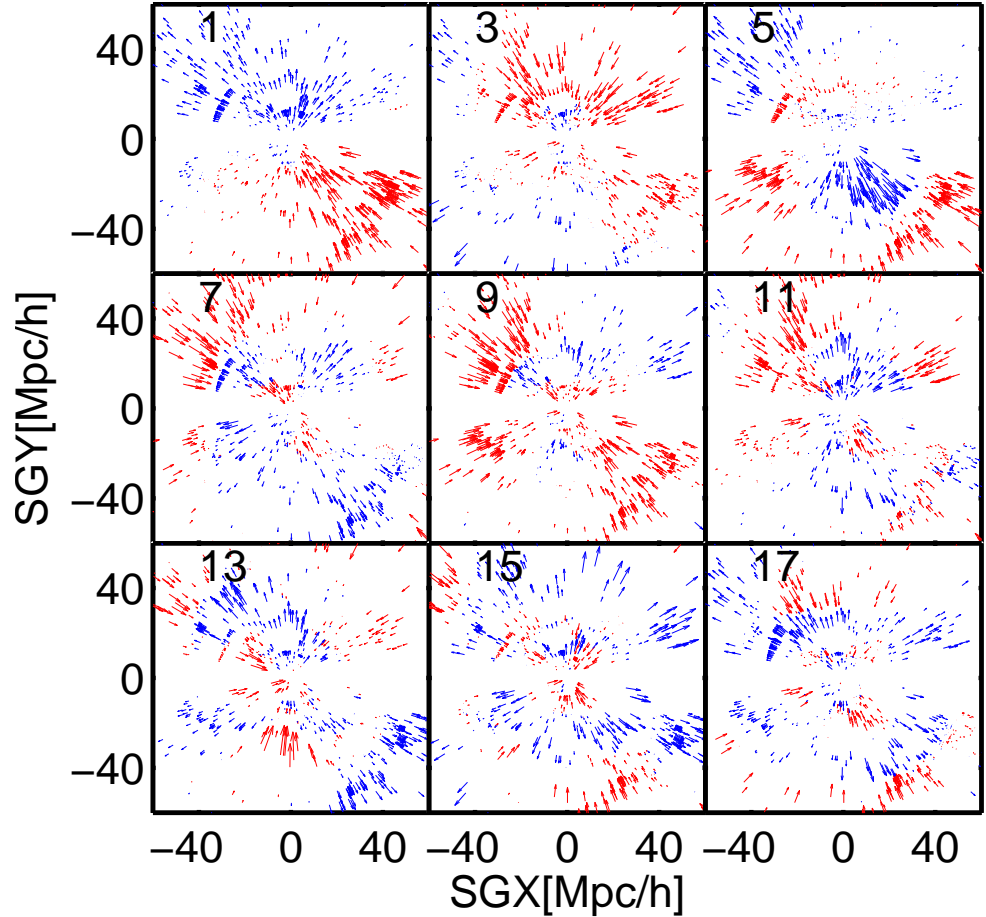}
\caption{ The  flow patterns of 9  modes for galaxies within $ 5\hmpc$ of the supergalactic plane.  The order of the mode is indicated
in the corresponding panel.  }
\label{fig:modes}
\end{figure*}

 The flow patterns of 
 9 of the modes are shown, respectively, in the 9  panels in figure (\ref{fig:modes}). The color scheme throughout this paper is normal: red (blue)  means outflowing (infalling), from the central point..
 We have extensively tested a broad variety of  choices for the the first mode. 
None of the  results of the analysis reported are sensitive to this choice of the 
 first mode. 
The figure shows that  the higher  order modes exhibit smaller scale structures. 
This is a direct result of the orthogonolization processes. The $j^{th}+1$ mode has to be orthogonal to the
all previous $j$ modes. In order to achieve that, the orthogonalized $j^{th}+1$ will pick more of the small scale structure.
 
\subsection{The ITF scatter and the number of modes}
\label{sec:TheITF}

Once the basis functions for the modal expansion are given, we proceed to
solve for the coefficients  $a^j$ in $P_i=\sum a^j F^j_i$   by minimizing $\chi^2_{\rm ITF}$ in (\ref{eq:chi}). 
The minimization is also performed, at the same time, with respect to the slope and zero point of the ITF. The  estimated slope is
$\gamma=-0.1297\pm 0.0015$ and $-0.13 \pm 0.0016$, respectively,   for 20 and 30 modes used in the flow model.
The raw slope before fitting the model is $\gamma=-0.1267\pm 0.0016$. The zero point plays no role at all here and 
we do not keep track of  its estimated values. 
All  estimated parameters, including $a^j$,  are independent of 
the (assumed constant) intrinsic scatter $\sigint$ in the ITF.  The velocity model can be used to estimate the unknown value of  $\sigint$. 
Given  the residual 
\begin{equation}
\Delta \eta_i=\eta_i -  (\gamma M_{0i} +\gamma P_i -\eta_0)
\end{equation}
we approximate $\sigint$ by  
\begin{equation}
\sigma_\eta^2=\sum_{i} (\Delta \eta_i)^2/N_{_{\rm d.o.f}} 
\end{equation}
where 
\begin{equation}
 N_{_{\rm d.o.f}}=N_{\rm g}-(j_{_{\rm m}}+2)
\end{equation}
  is the number of degrees of freedom taking into account 
that the minimization of $\chi^2_{\rm ITF}$ is done with respect to $j_{_{\rm m}}$ coefficients plus the slope and zero point of the ITF.
This $\sigma_\eta$ will  decreases as the number
 of modes, $j_{_{\rm m}}$,  in the expansion  is increased. 
If   $j_{_{\rm m}}$ is too large then the higher order modes will be dominated by noise.
If $j_{_{\rm m}} $ is too small then the model may miss significant components of the underlying 
true galaxy velocities.  

The optimal  range of $j_{_{\rm m}} $ for our comparison can be seen
by inspecting the behavior of $\sigma_\eta$ as a function of $j_{_{\rm m}}$. 
 The (blue)  circles in the top panel of figure \ref{fig:chi_TF} shows $\sigma_\eta $ versus the number of modes 
 for modes generated from the seed field $\Vs=\vg(\beta=0.2) $ (see \S\ref{sec:refine}). 
 Most of the reduction in $\sigma^2_\eta$ is already achieved  the first mode. 
  This is very encouraging since this means that $\vg$  picks up a significant 
  contribution of the velocities  as described by the ITF data. It also means that both the 2MRS and the ITF data are likely to provide approximations to the underlying flow  field. 
  However,  the 2MRS predicted field $\vg$ deviates from the underlying field by the presence of correlated  errors in the 
  reconstruction scheme. 
The inclusion of  additional  expansion modes in the ITF velocity model will dissolve these
  errors. The average  reduction in the variance $\sigma^2_\eta$ per mode becomes insignificant beyond $j_{_{\rm m}}=64$; 
  the average reduction per mode for the first 64 modes is 10 times larger than that for the  next  75 modes. 
  An F-test also confirms that the reduction in the variance marginal  beyond $ j_{_{\rm m}}=64$.
  Hence we will approximate $\ssigint=0.0558$, the value acquired by  $\sigma_\eta$  
  for 64 modes in the velocity model.  Therefore, the F-test argues that $64$ is the maximum number modes needed to 
  model the ITF. 

  The next step is to determine the minimum number of modes needed to describe the ITF flow assuming  that 
  $\sigint=0.0558$. To do so we tabulate  $\chi^2_{\rm ITF}$  a function of $j_{_{\rm m}}$ and 
 compute the probability $Q=Q(\chi^2_{\rm ITF}|N_{_{\rm dof}})$  that the value $  \chi^2_{\rm ITF} $ is exceeded by
chance \citep[c.f.  \S6.2 in ][]{press}.
The values of  of  $\chi^2_{\rm ITF}/N_{_{\rm dof}}$ and Q are represented  as the (blue) circles, respectively,  in the middle and bottom panels of figure \ref{fig:chi_TF}. 
For $j_{_{\rm m}}=64$, we get $\chi^2_{\rm ITF}/N_{_{\rm d.o.f}}=1$  corresponding to  $Q=0.5$, in accordance with our choice of  $\sigint$.
Without a velocity model, i.e. $j_{_{\rm m}}=0$, we get $(\chi^2_{\rm ITF}/N_{_{\rm dof}}, Q) =(1.125, 3\times 10^{-6})$. This 
exceedingly  low $Q$  
rejects a vanishing velocity field  with very high confidence level (CL).
Including the first mode alone gives  a highly significant improvement: $(\chi^2_{\rm ITF}/N_{_{\rm d.o.f}},Q)=(1.031, 0.12)$. 
The hypothesis that $ \chi^2_{\rm ITF}$ value  corresponding to the  first mode is obtained by chance is 
rejected only at the 0.12 CL. This is encouraging since the first mode velocity field is proportional to the
2MRS predicted velocities, $\vg$ (for $\beta=0.2$).
 For $j_{_{\rm m}}=30$ and   20  we get $(\chi^2_{\rm ITF}/N_{_{\rm d.o.f}},Q)=(1.01, 0.29)$ and  $ (1.014, 0.2)$,   respectively.

For comparison with $\Vs=\vg$, the corresponding results for  random choice of the seed field, $\Vs$, are shown as the (red)  crosses  in the top and bottom panels of figure \ref{fig:chi_TF}.
With a random $ \Vs$, about 20 modes are
needed to reduce $\sigma_\eta$ to the level achieved by the single mode $\vg$.   
 
\begin{figure*} 
\centering
\includegraphics[ scale=0.8 ,angle=00]{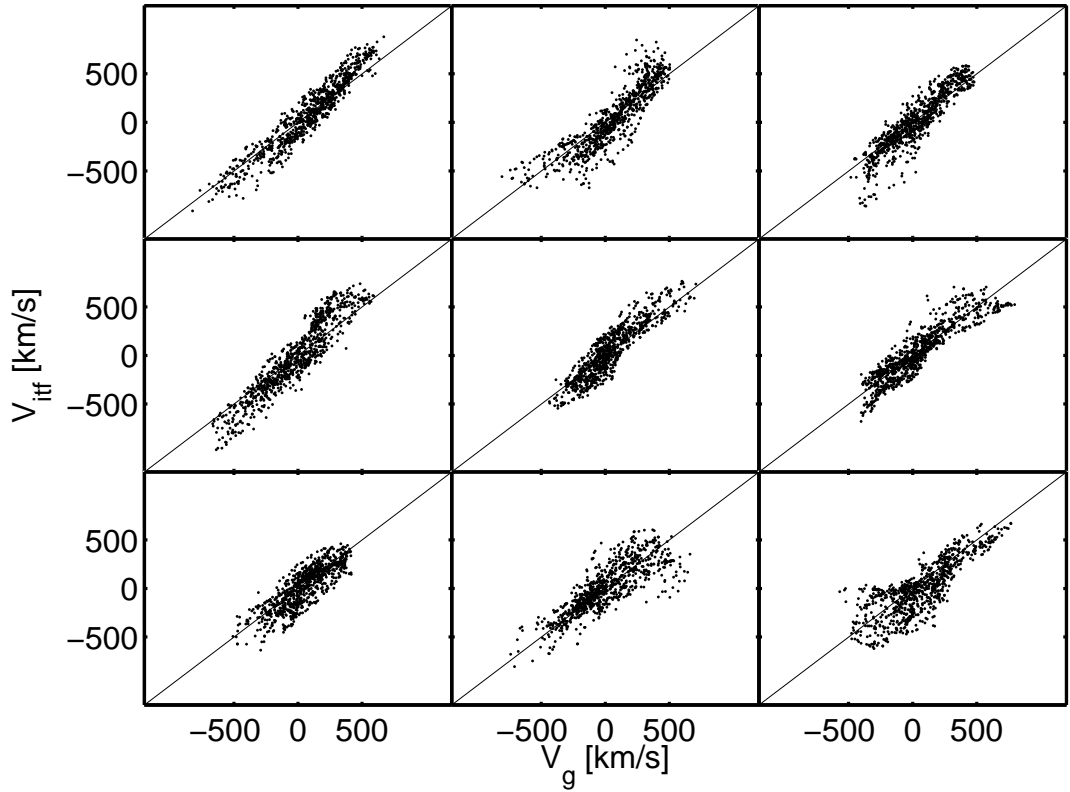}
\caption{ Scatter plots of $\vitf$ versus $\vg$ (expanded in 20 modes) for galaxies in 9 mock catalogs. 
About 800 galaxies are plotted for each mock.}
\label{fig:mocks}
\end{figure*}

\begin{figure} 
\centering
\includegraphics[ scale=0.8,angle=00]{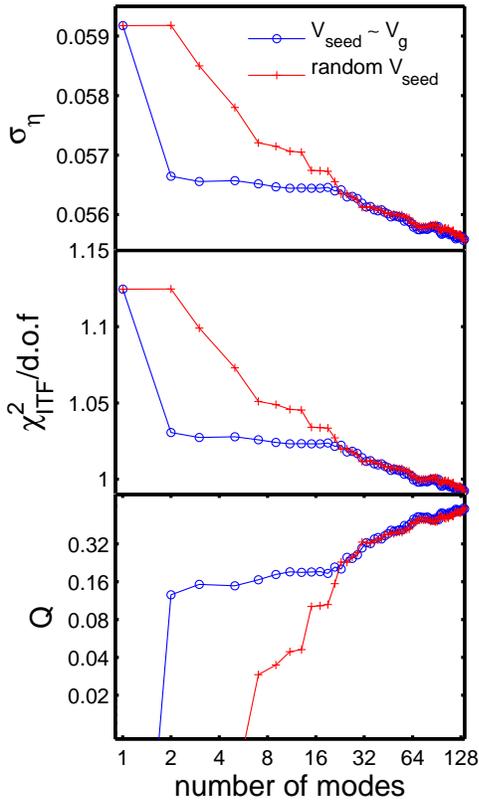}
\caption{ {\it Top:} The rms value,  $\sigma_\eta$, of the ITF   as a function of the number of modes used 
in the velocity model.   {\it middle:} $\chi^2_{\rm ITF}$ per degrees-of-freedom versus the number of modes. 
It is unity at 64 modes. 
{\it Bottom:}  The probability that the 
$\chi^2_{\rm ITF}$ exceeds a certain value by chance,  as a function of the number of modes. The chi-square is computed assuming an intrinsic scatter  $\sigint=0.0558$. The value $Q=0.5$ is achieved at 64 modes. (see text for details)}
\label{fig:chi_TF}
\end{figure}

\begin{figure}
\centering
\includegraphics[ scale=.8 ,angle=00]{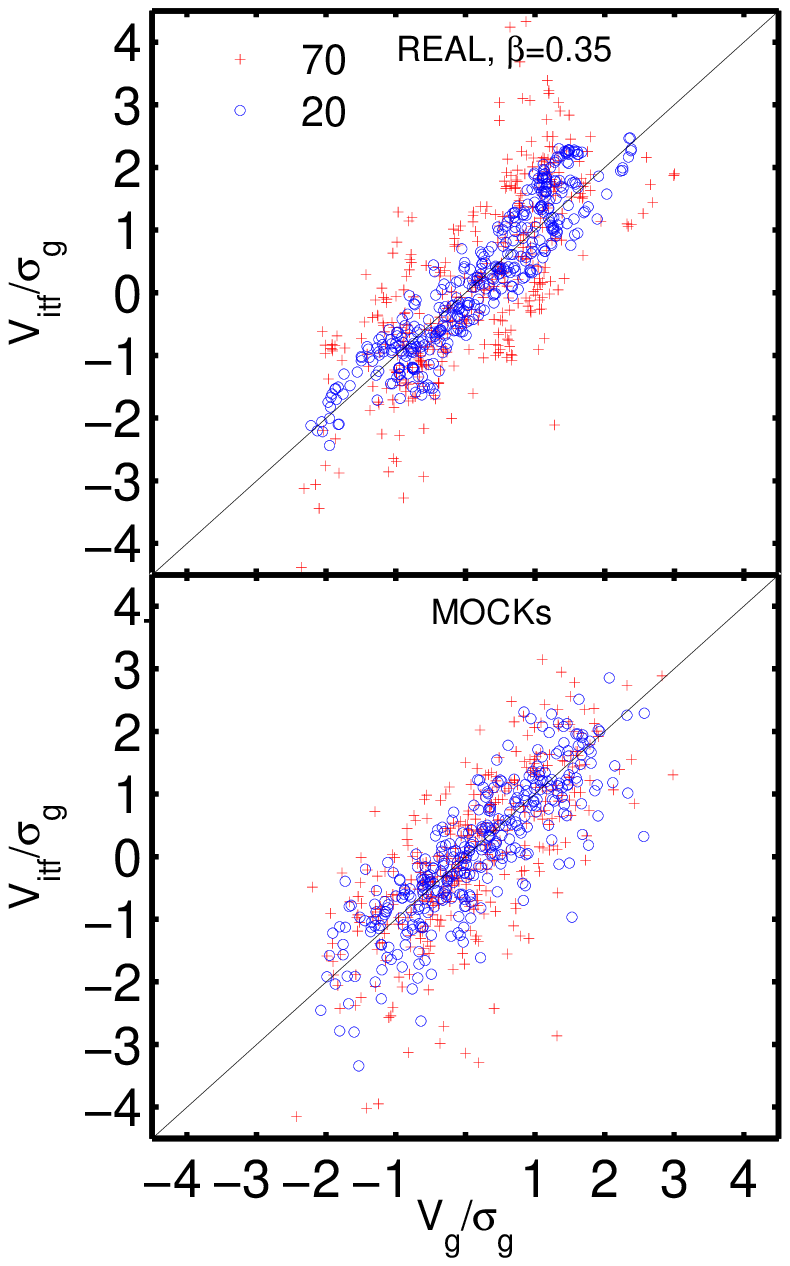}
\caption{ The peculiar velocities  $ \vitf$  versus $\vg$ of galaxies in  the real data (top) and the mocks (bottom)
for 20 and 70 expansion modes, as indicated in the figure.
 The real $\vg$  has been reconstructed with $\beta=0.35$ and $\alpha=0$.
Velocities of  about 400 randomly selected galaxies are plotted in each panel where each  mocks is represented by about   25 galaxies.  All velocities in the top panel are normalized by the rms value,  $\sigma_g=233\kms$ of $\vg$, while velocities 
in the bottom panel are normalized by the 
 rms value of $\vg$ of  their corresponding catalogs. }
\label{fig:vv}
\end{figure}



\subsection{What is the purpose of the ITF machinery?}
The expansion of the gravity field is conceptually very clean when computed in the LG frame (ND95). 
The Poisson-like equation for the 3-D gravitational field has been solved as a sum over the spherical harmonic functions $Y_{lm}$ times 1-D functions of $r$ that satisfy physically reasonable boundary conditions at the origin.  For the purposes of the ITF solution, we furthermore quantize the radial solutions with quantum number $n$.   

The ITF method is backward from the usual methodology of TF applications; one does not fit  curves to the scatter of peculiar velocities. Instead, the $\chi^2$ equation \ref{eq:chi} is minimized by the addition of linear combinations of the orthonormal functions of  $n,l,m$,  where each describes a set of large-scale flow that satisfy the boundary conditions.   Furthermore, we have endeavored to form a first, 'seed',  mode based on linear growth rate, but in which large scale graininess is filtered out by the use of figure \ref{fig:Rs}.  

The individual galaxy's peculiar velocity  enters by equation \ref{eq:P_i}, with differential 
$d P_i    \propto d u_i /z_i $, and since the uncertainty of peculiar velocity $u_i$ is proportional to redshift, the uncertainty of $P_i $ is redshift independent.  This means that each object is given equal weight in a fit, and our window function is therefore equivalent to the display of figure \ref{fig:vitf}, which shows the positions of the SFI++ galaxies.

 
\section{Mock catalogs and error analysis}
\label{sec:mocks}
As a  measure of the agreement between the TF  and predicted velocities by means of  the corresponding 
expansion coefficients, $\aitf$ and $\ag$, we will consider the 
$\chi^2$ function
\begin{equation}
\label{eq:chib}
\chi^2=\sum_{(j,k)=1}^{j_{_{\rm m}}} \left(\ag^j-\aitf^j \right)\left(\sigma_a^2+\xi_{_{\rm g}}\right)^{-1}_{j,k}\left(\ag^{k}-\aitf^{k} \right)
\; .
\end{equation} 
The parameters $\alpha $ and $\beta$ will be obtained by minimization of this function. 
The covariance of the residual, $\aitf-\ag$,  is 
the sum of the covariance matrices of the errors in the estimation of  $\aitf$ and 
$\ag$, respectively.
Thanks to the orthonormality of the basis functions, the  error covariance in  the estimation of $\aitf$ is diagonal with constant terms
$\sigma_a^2=\left(\sigma_\eta/\gamma\right)^2$ (see \S\ref{sec:itfd}  and DNW96).
The  matrix $\xi_{_{\rm g}}$  represents  the covariance of the errors in the 
determination of  $\ag$. 
The origin of these errors is as follows:
\begin{enumerate}
\item {  Equation \ref{eq:phis} is expected to predict  reliable velocity fields 
 only for small amplitude fluctuations. 
Small scale nonlinear deviations from linear theory inevitably leak to large scales.} 
\item{ The  2MRS is a finite number  sampling of the underlying density field. 
This leads to ``shot-noise" errors in the estimation of the density field.}
\item{ Small scale random motions of galaxies, especially in  groups and 
clusters, give rise to a smearing of the distribution of galaxies along the line of sight 
in redshift space.}
\item{ There is a possible large scale stochastic biasing \citep{dh99,sb00,wpl05} between the galaxy distribution and the mass fluctuations.} 
\end{enumerate}

The only way to achieve a reliable estimate of  $\xi_{_{\rm g}}$ taking into 
account all of these complicated errors is by means of mock catalogs designed to 
match the general properties of the 2MRS. 
A parent simulated catalog of the whole 2MASS catalog has already been
prepared \citep{delucia} by incorporating semi-analytic galaxy formation models
in the Millennium simulation \citep{mill}. 
From this parent catalog we have drawn 15 independent mock 2MRS  catalogues 
satisfying the following conditions:
\begin{enumerate}
\item{ The ``observer" in each mock is selected to reside in a galaxy with 
a quiet velocity field within $500$ km/s,  similar to the observed universe.  That is, the  central server sees only one cluster  that has high enough peculiar velocities to result in negative redshifts.   Recall that in the LG  frame, the only galaxies with negative redshift are in the Virgo cluster. }
\item{  The motion of the central galaxy is 500 to 700 km/s. } 
\item{ The density in the environment of the local group, averaged over a sphere of 400 km/s radius ,  is  less than twice the normal. }
\end{enumerate}

Corresponding mock ITF catalogs were also prepared.
A counts-in-cells statistics shows that the distribution of galaxies in the mocks is 
 unbiased relative to the dark matter, i.e. $b=1$.
 
The preparation of the mocks for velocity  reconstruction is done in the 
same way as the real data.
Equation (\ref{eq:phis}) is used to generate prediction of mock $\vg$ with $\beta=f(\Omega,\Lambda)/b =0.47$ corresponding to  $b=1$ and 
$\Omega=0.25$ and $\Lambda=0.75$ 
as in the Millennium simulation.
The mean  of the rms values of $v_g$ in the mocks is $269\kms$ and the standard deviation from 
this mean is $56 \kms$.
For contrast, the rms value of $v_g$ derived from the 2MRS with $\beta=0.35$ is
$233 \kms$.  

To better illustrate the covariance between the residuals in the mocks, we plot in 
 figure \ref{fig:mocks}  the velocities  $\vitf$ versus $\vg$ for 9 individual mocks.  These are velocities expanded with 20 modes. 
 Note the similarity between the structure of the distribution of points in the individual panels and 
 the top panel in figure \ref{fig:vv}, showing $\vitf$ versus $\vg$ (with $\beta=0.35$) for the real data.

\subsection{The error covariance matrix} 
The covariance matrix $\xi_{_{\rm g}}$ is computed from the 15 mocks  by 
projecting the correlation function, $\xi_{_{\rm P}}$, of the residuals
$\Delta P=P_{\rm itf} -P_{\rm g}$ onto the basis functions, where 
 $P_{\rm itf}=5{\rm  log}(1-\vitf/cz)$ and correspondingly 
for $P_{\rm g}$. That is 
\begin{equation}
\label{eq:xiaa}
\xi_{_{\rm g}}(j,k)=<(\aitf^j-\ag^j)(\aitf^k-\ag^k)>=\sum_{i,i'} F^j_i \xi_{_{\rm P}}(i,i') F^k_{i'} \; ,
\end{equation}
where the summation in the last term on  the right-hand side is over all data galaxies and 
$F$ are the basis  functions used for the real data.  
In this calculation,    the $\vitf$ velocities are  reconstructed from an ITF relation without adding the internal scatter of the TF relation.
The reason  is that the error in $\aitf$ resulting from the intrinsic scatter has a simple analytic  form
given by $\sigma_a$.   
The function $\xi_{_{\rm P}}$ is computed from the 15 mocks as follows. 
Denote   line-of-sight and projected separations in redshift space by, $s_\parallel$ and $s_\perp$, respectively.
For each mock we tabulate the average $ <\Delta P_1\Delta P_2> $ over pairs 
with separations defined by the grid.    
We then normalize this quantity by the variance of 
$\vitf$ (for zero ITF instrinsic scatter) in the corresponding mock. 
This  is reasonable since the  rms values of the velocity field vary considerably among the  mocks and some of them 
are significantly different than the real data. To minimize this cosmic variance and to derive $\xi_{_{\rm g}} $ given the observed 
rms value of the velocity, this normalization of $ <\Delta P_1\Delta P_2> $ for each mock is prudent.
The  average over all mocks is then computed and interpolated from the grid onto the 
actual pair  separations in the TF catalog to obtain   the normalized $\xi_{_{\rm P}}$.
The normalized covariance matrix $\xi_{_{\rm g}}$ is then computed according to   (\ref{eq:xiaa}) and 
scaled  by a factor matching the velocity variance estimated from the observed $\vitf$. 

\section{The reconstructed velocities}
\label{sec:results}
This section presents 
a visual inspection of the fields, and assess the coherence of the residual  $\vitf - \vg$   by means of 
a velocity correlation analysis. 
The quantification of the agreement between the fields and the extraction constraints on $\alpha$ and $\beta$  will
be deferred to later sections.

\begin{figure*}
\centering
\includegraphics[ scale=0.6,angle=00]{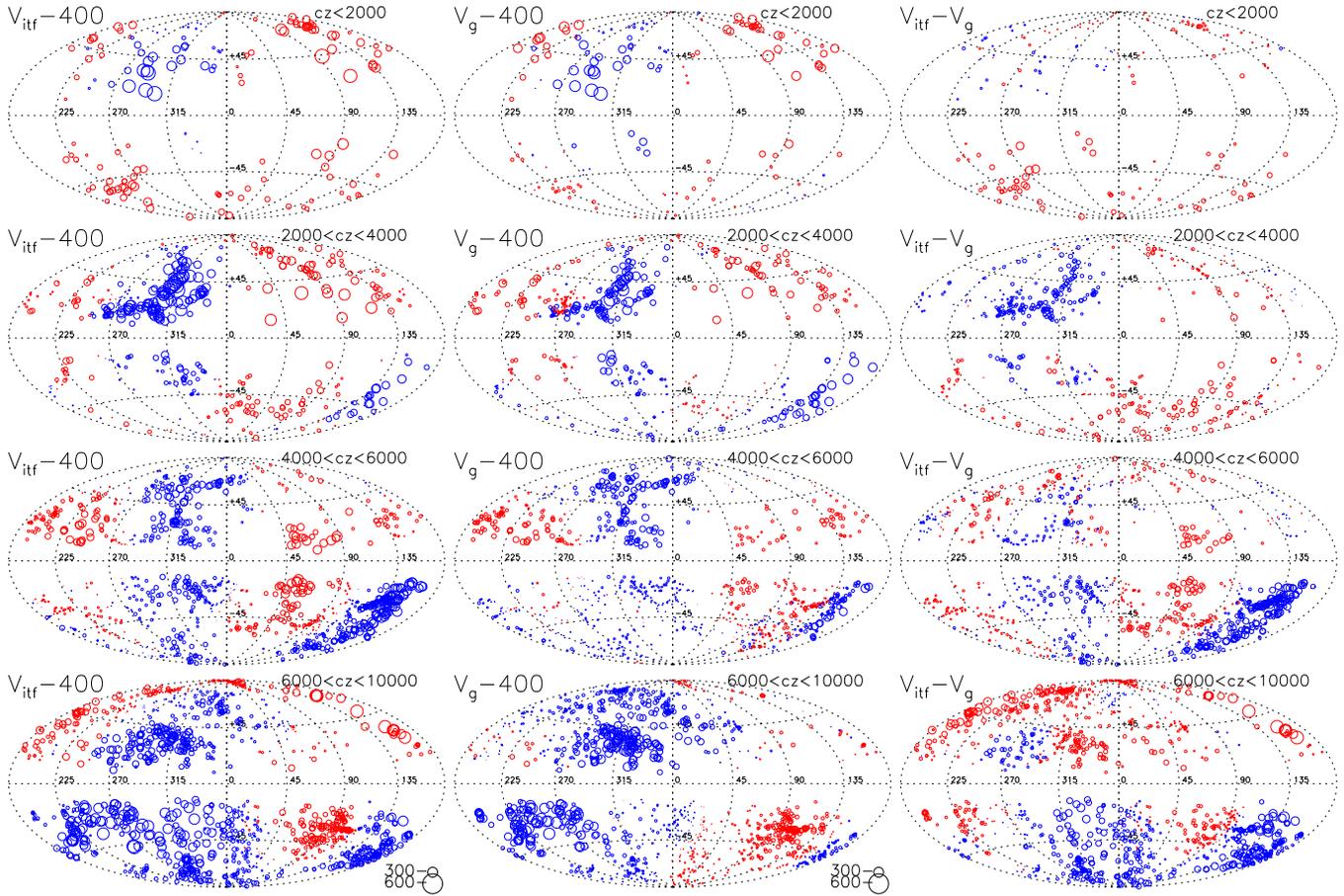}
\vspace{0pt}
\caption{  The derived peculiar velocities $\vitf$, $\vg$, and $\vitf -\vg$  of galaxies on aitoff projections on the sky in galactic 
coordinates. The rows correspond to galaxies with $cz<2000$, $2000<cz<4000$ 
$4000<cz<6000\kms$ and $6000<cz<10000$ km/s, respectively. The size  of the symbols is linearly proportional to the velocity amplitude (see key to the size of the symbols given at the bottom of the figure). In order to better see the differences,  a 400 km/s dipole, in the direction of the CMB dipole, has been subtracted from the $ \vitf$ and $\vg$ velocities.  Note that $\vitf -\vg$ is considerably smaller than $\vitf $ or $\vg$, even for the most distant galaxies.}
\label{fig:vitf}
\end{figure*}

\begin{figure*}
\centering
\includegraphics[ scale=.65 ,angle=90]{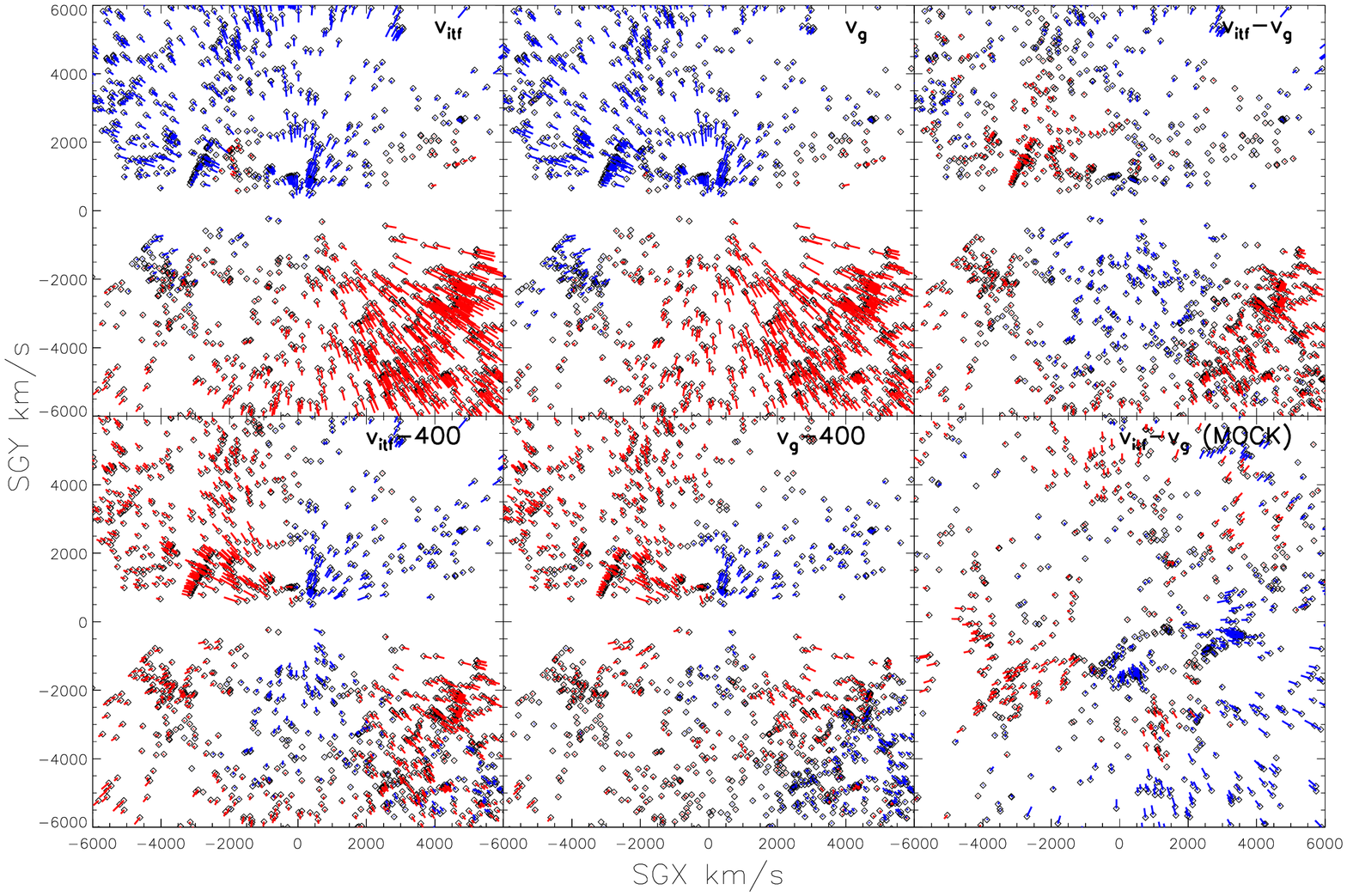}
\vspace{10pt}
\caption{Supergalactic plane projection, $|SGB| <30$,  of the derived flows.  To better see the differences in the plots, a dipole of 400 km/s towards the CMB pole has been subtracted from the fields, and is shown in the bottom left and bottom center.  A quadruple velocity is now visible in the plots.  The points are drawn at the estimated distance of an SFI++ galaxy, and the line, blue or red, is drawn to the galaxy's redshift. In other words, the length of the arrow is the peculiar velocity. The lower right plot shows  $\vitf-\vg$ for a mock catalog, and the upper right shows $\vitf-\vg$ for the data.  They have very similar degrees of coherence.} 
\label{fig:6fig}
\end{figure*}

\subsection{Visual inspection of the flows}

Blue dots and red crosses of figure \ref{fig:vv} show velocities expanded in 20 and 70 modes, respectively.  
In the bottom panel,  $\vitf$ versus $\vg$ from the mocks are shown. In mock $\vitf$ velocities are obtained from a fake ITF relation with 
an intrinsic slope $\gamma=-0.1$ and a scatter with  $\sigma_\eta=0.05$.
For the sake of clarity only 400 randomly selected  galaxies are shown in either panel. 
Further, each mock is represented by about 25 galaxies (randomly selected). 
The velocities are scaled by the corresponding 
rms value of $\vg$ in the corresponding catalog.

There is a  an excellent  overall agreement  between $\vitf$ and $\vg$ for 20 modes, both in the real data and the mocks.  A good agreement  prevails  even for 70 modes despite the clear enhanced noise  contamination.
 In the real data, the rms of $\vitf-\vg$   for 20 expansion modes is $99\kms$, significantly smaller than $\sigma_g$. 
For 70 modes the rms values of   $\vg$ and $\vitf-\vg$ are $238\kms $ 
and $231\kms$, respectively. Both panels  show clear structures in the 
distribution of points, implying strong covariance between the residuals, $\vitf -\vg$, in the real, or mock, data.
Because the bottom panel  represents 
 random selections of galaxies from all the  mocks, the covariance pattern between the velocities is diluted in the distribution of points. The covariance pattern is, however, clear in figure \ref{fig:mocks} where scatter velocity plots for a few mocks are shown individually. 

In the aitoff projections in Figure \ref{fig:vitf} we plot the TF peculiar velocities, $\vitf$ and the derived gravity modes, $\vg$, for galaxies 
in redshift shells, $cz<2000$, $2000<cz<4000$, $4000<cz<6000$, and $6000<cz<10000\kms$.  The projections are in galactic coordinates centered on $l,b=0$ and with $b=90$ at the top.  
Figures \ref{fig:vitf} and  \ref{fig:6fig}  show $\vg$ with $\beta=0.35$.  
The  rightmost plots are the residuals $\vitf-\vg$.  The key point is to note that the residuals are small for the entire sky and have amplitude that is constant with redshift.  
  The amplitude and coherence of the residuals $\vitf-\vg$ is the same as for the mock catalogs 
  in figure \ref{fig:6fig},
 where for example the lower right picture shows $\vitf-\vg$ for a mock catalog. It is not very dissimilar  from the real plot of $\vitf-\vg$ in the upper right, demonstrating the feasibility of the entire method.
 
 Note the quadrupole pattern for $cz<4000\kms$ in figures \ref{fig:vitf} and  \ref{fig:6fig}, visible after $400\kms$ has been subtracted from the flow. This has been previously noted by \cite{hht07} on the basis of the flows detected in 133 SNe.  The quadrupole is the typical pattern observed in nbody simulations and is the principle mode of collapse to a 1-D structure. 

There is amazingly overall good agreement between the large scale motions 
as described by $\vitf$ and $\vg$. The  residual  velocities  are coherent over large scales but they are clearly of smaller amplitude  than $\vitf$ and $\vg$. 
Note that residuals shown in $\vitf-\vg$, particularly visible in the shell $6000<cz<10,000\kms$, are dominated by $l=4$, because the fit for the reduction of the TF $\chi^2$ is limited at $l=3$ modes.

\begin{figure}
\centering
\includegraphics[ scale=0.5,angle=00]{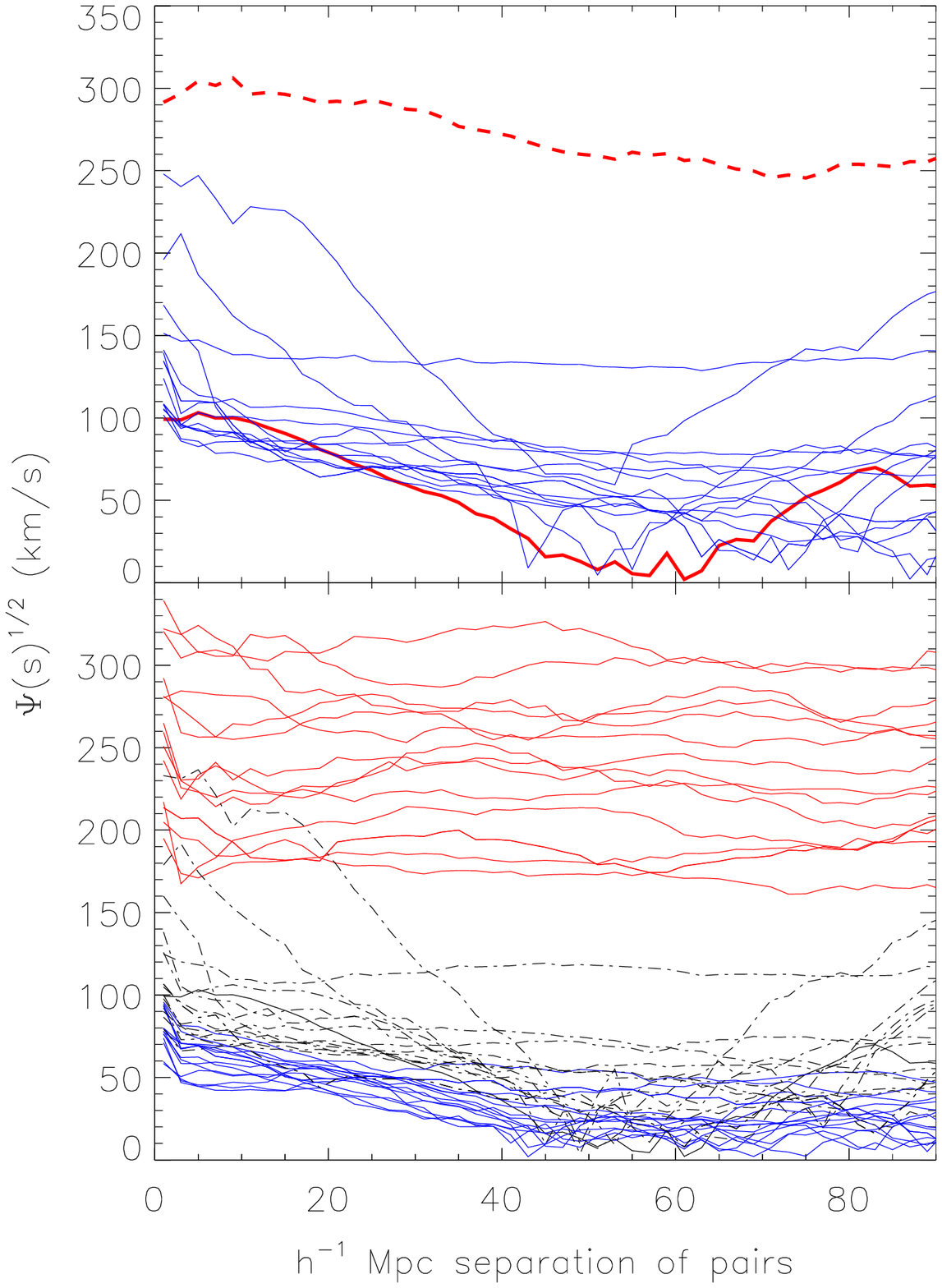}
\vspace{10pt}
\caption{{\it Top: } The velocity correlation of the real data and 15 mock catalogs. 
The dashed red and solid red curves curve are, respectively,   the correlations of $\vitf$ and $\vitf - \vg$ in  the real data.
  The blue lines are each correlations of $\vitf-\vg$ for the mock catalogs.  {\it Bottom:}  Velocity correlations  for 15 mock catalogs.  The red curves are
   the velocity of $v_{itf}$, the dot-dashed curves show the correlation of $(v_{true}-v_{g})$, and the blue curves correspond to  $v_{true}-v_{itf})$.  
Both $v_{true}$ and $v_{g}$ are first smoothed with the 20 mode expansion before the autocovariance is computed. Note that the correlation of $\vitf-\vg$ is only slightly worse than the correlation of $v_{true}-v_{gs}$, showing that the velocity reconstruction dominates the errors.  Note also that we are plotting the square root of the velocity correlation $\Psi$.  }
\label{fig:psiv}
\end{figure}

\subsection{Correlations} 
The residuals, both in the real and mock data, have error fields, $ \vitf - \vg$,  that show large regions of coherence.   To address the significance of these errors, we show in figure \ref{fig:psiv} the velocity correlation function, \citep{gd89}, defined as 
\begin{equation}
\label{eq:Psiv}
\Psi(s; {u})=\frac{\sum_{pairs}  u_1u_2 {\rm cos}\theta_{12}}{\sum_{pairs} {\rm cos}^2\theta_{12}}
\end{equation} 
where  the sum is over all pairs, 1 and 2, separated by vector distance $\bf s_{12}$  
(in redshift space),  $ {\rm \theta_{12}}$ is the angle between points $1$ and $2$,  and $u$ is either $\vitf$ 
(dashed red) 
or $\vitf - \vg$ (red for data, blue for 15 mock catalogs),    
At small lags for the real data, the function $\Psi(r; {\vitf-\vg})$ is  a factor of 3 less than 
$\Psi(s;{ \vitf})$, about the same as for the mock catalogs. Note how the large coherence of $\vitf$ is enormously diminished in 
$\Psi(s>2000 {\rm km/s};{\vitf-\vg})$. This shows that the coherence seen in the residual field, 
figure \ref{fig:psiv}, is expected and is not a problem. The large scale drift of a sample is demonstrated by the persistent amplitude of $\Psi$ beyond $\approx 60-80$ Mpc.

The bottom panel of figure \ref{fig:psiv} shows velocity correlations for 15 mock catalogs where the actual velocity, $v_{true}$, generated in the nbody code and then smoothed with the 20 mode expansion,  can be compared to either $\vitf$ or $\vg$.   Note that the raw velocities, $v_{itf}$ (red), have enormous correlation that reaches large lag, while the correlations, $(v_{true}-\vitf)$, (blue) are extremely small.  This is because the only difference with $v_{true}$ is the gaussian error in $\Delta\eta = .05$ that affects $\vitf$.  The blue curves show this error is not a problem, because the mode expansions are insensitive to gaussian noise in the 2500 galaxies, i.e. they are essentially perfect.  This demonstrates that even though the TF noise is as large as for the actual data, the ability to find the correct flow, when characterized by only 20 numbers, is intact.  

Note also that the auto-covariance of $(v_{true} - \vg)$ (dot-dashed curves) is also is greatly reduced from that of $\vitf$. Recall that $\vg$  assumes linear theory estimated from the distribution of ~20000 galaxies. Occasionally the correlations are badly mistaken, when a large cluster (much larger than Virgo) is in the foreground and complicates the difference between physical and redshift space separations, but $\vg$ is always an excellent approximation to the TF velocity.

\begin{figure} 
\centering
\includegraphics[ scale=0.78 ,angle=00]{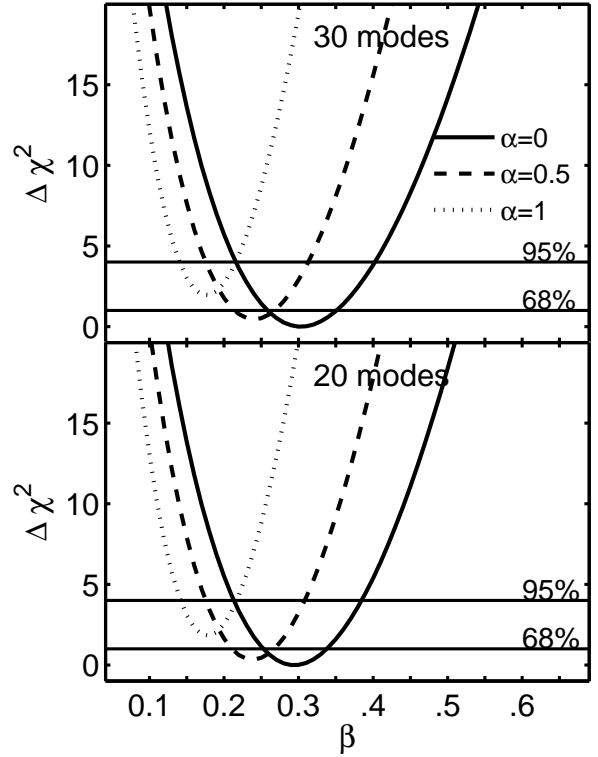}
\caption{The difference $\Delta \chi^2=\chi^2-\chi^2_{\rm min} $ versus $\beta$ computed for three values of $\alpha$, as
indicated in the figure. Horizontal lines mark the 68\% and 95.4 \% CLs. }
\label{fig:chibeta}
\end{figure}

\begin{figure}
\centering
\includegraphics[ scale=0.5,angle=00]{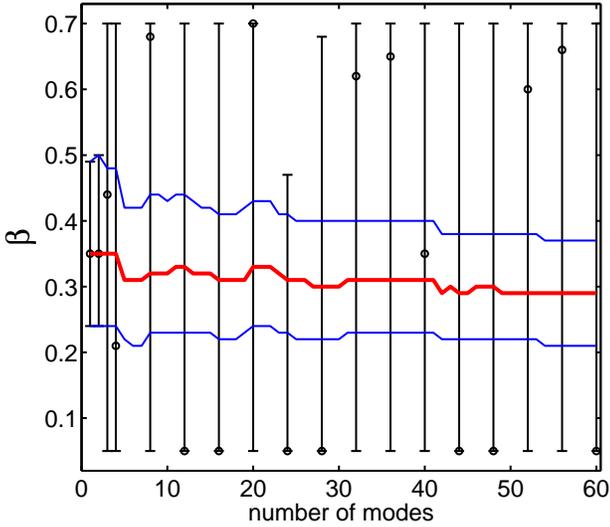}
\caption{The thick  solid red line is  best fit $\beta$ as a function of the number of modes in the velocity model, for $\alpha=0$. 
The thin blue lines mark the $95\% $ CL on the best fit. 
The circles denote the ``differential"  best fit  $\beta$ obtained with the  single $j^{th}$ mode. After the first 4 modes, 
only 1 in 4 modes are 
represented.  The errorbars attached to the circles are 
correspond  to $95\%$   CLs on the differential best fit. }
\label{fig:betaj}
\end{figure}
\begin{figure} 
\centering
\includegraphics[ scale=0.4 ,angle=00]{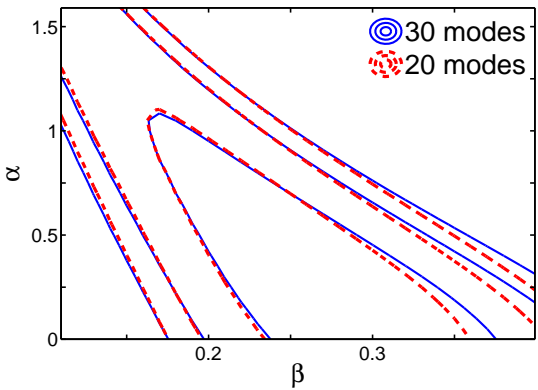}
\caption{ Contour plot of  $\Delta \chi^2$ in the plane of $\alpha$ and $\beta$.  The contours are 2.3, 6.17 and  9.2 corresponding to  CLs of 68\% 95.4\% and 99\%, respectively. }
\label{fig:chiab}
\end{figure}

\section{The Constraints on $\alpha$, $\beta$}
\label{sec:const}
Equipped with the error covariance matrices, we proceed to 
minimize $\chi^2$ in equation (\ref{eq:chib}) with respect to $\alpha$ and $\beta$. We shall present 
detailed results for $\alpha$ and $\beta$ for fields expanded in 20 modes and 30 modes. 


The minimization is done by computing $\chi^2$ on a grid of values in the plane $\alpha$ and $\beta$.
At the minimum  point  $\chi^2=\chi^2_{\rm min}=21.5$ which is very reasonable given that 
the standard deviation from the expected value of 22 (20 mode coefficients plus TF slope and zero point)  is $\sim \sqrt{44}\approx 7$ \citep{press}. 

Figure \ref{fig:chibeta} shows the difference $\Delta \chi^2=\chi^2 -\chi^2_{\rm min}$ as a function
of $\beta$ for three fixed values of $\alpha$, as indicated in the figure.
The horizontal lines indicate 68\% and 95\% confidence levels (CLs).
Figure \ref{fig:chiab} is a contour plot of $\Delta \chi^2$ in the plane of $\alpha$ and $\beta$.

The higher frequency  basis functions should probe smaller scales.
Hence, if our assumption of linear bias which is independent of scale is valid then 
varying the number of modes, $j_{_{\rm m}}$, should yield consistent constraints on  $\beta$. 
Figure \ref{fig:betaj} shows the best fit $\beta$ (thick red curve) and the corresponding $95\%$ (2$\sigma$) CLs (thin blue solid lines) versus the number modes in the expansion, for $\alpha=0$.
The circles show the ``differential" best fit $\beta$ obtained from a single mode as a function of the order of 
the mode. The $2\sigma $ errorbars on this differential $\beta$ are significantly enhanced  beyond the 
second order mode.  A few points lie at the ends of the errorbars corresponding to best fit $\beta$ 
obtained at either $0.05$ or $0.7$ which are the bounds  of the range of $\beta$ values used 
in the 2MRS reconstruction. 
 There is a hint that  $\beta$ declines with increasing $j_{_{\rm m}}$ but this is completely dominated by the 
 noise. 
 This figure shows clearly that we get consistent constraints on $\beta$ (the red curve) when varying the 
 number of modes in the expansion. Further, it shows that most of the signal is contained 
 the very few first modes.

\section{Discussion}
\label{sec:disc}
The  analysis reported demonstrates a good match between  the 2MRS predicted and TF observed velocities. 
The analysis is unique in several  respects.
Firstly, it completely avoids dealing with covariance matrices of errors in the 
velocities estimated from the TF sample, 
Secondly,  it uses   elaborate mock galaxy catalogues to compute the error covariance in the predicted velocities from the redshift survey.
   Thirdly, the  TF and predicted velocities are filtered in a   very similar fashion, taking special care  to minimize the effects of noise in the comparison.  
   In this analysis faint galaxies with $M>-20 $ are excised from the TF catalog since they systematically show strong deviations from the linear TF relation. 

The comparison yields $\beta=f/b=0.33\pm 0.04$ ($1\sigma$ error). 
The quoted error is not actually far from the  limit of what the current data can constrain in the 
absence of any errors on the 2MRS predicted velocities
\footnote{In the absence of gravity errors and  for $\beta$ close to the best fit value $\beta_{_{\rm 0}}$, the $\chi^2$ function is approximated as
\[ \chi^2= \sum_j[\aitf^j -B(\beta)  a_{_{\rm g 0}}^j ]^2/\sigma_a^2\] where  $a_{_{\rm g 0}}^j$ correspond to reconstruction 
with $\beta=\beta_{_{\rm 0}}$ and $B=(\beta/\beta_{_{\rm 0}})(1+1.5\beta_{_{\rm 0}})/(1+1.5\beta)$ approximates the
dependence of $\vg$ on $\beta$ (in contrast to the dependence $\beta/\beta_{_{\rm 0}}$ in reconstruction from galaxy distribution in real space). 
The $1\sigma $ error in  $B$ is $\sigma_{\rm a}/\sqrt{\sum_j (a_{_{\rm g 0}}^j)^2}\approx 0.06$, where we took $\sigma_{\rm a}=\sigma_\eta/\gamma=0.43$ and  $ \sum_j (a_{_{\rm g 0}}^j)^2=52 $ as given from the solution with $\beta=0.35$. This error in $B$ translates into 
an error of $0.03$ in $\beta$ which is close to the error obtained with the full analysis.}.
A moderate reduction of the errors  by 
a factor of two requires a significant enlargement in the number of peculiar velocity measurements by a factor of 4, which 
could be done if dedicated time is available, but the TF samples already use the best local galaxies. 
Going to larger distance is not the answer, as the error of a peculiar velocity increases linearly with the distance, and the 2MRS density field becomes very dilutely sampled.  Two surveys, WALLABY, to be undertaken by the ASKAP telescope  in Western Australia, and ALFALFA, an ongoing project at Arecibo, will hopefully produce good TF data
for $cz<12000$ km/s.
Another strategy would involve peculiar velocities inferred  from
more precise distance indicators than the TF relation. SNe and surface brightness fluctuations 
techniques are likely candidates, but such measurements are available for a much smaller number of galaxies.  
Larger samples of local SNe are turning out to have increased errors (Ganeshalingam, private communications), significantly larger than previously measured \citep[e.g.][]{r97}.

The good match between the gravity and velocity  fields  implies  that they probe the same underlying 
potential field within the framework of the gravitational instability paradigm for structure formation.
 The agreement is achieved assuming a  linear  biasing relation between mass and galaxies on 
large scales. Linear biasing  is consistent with theoretical predictions \citep[e.g.][]{kn97} for the large scale clustering of galaxies. Further, no scale-dependent biasing seems to be required by the velocity comparison.
However, the theoretically expected scale dependence of the bias factor \citep[e.g.][]{pesm,dej10} is well below
 the level which can be probed by the velocity comparison carried out here.
There is also no clear indication for a scale dependence bias  from the observed 
galaxy clustering on  the relevant scales  \citep[e.g.][]{ver2df}.

DNW96, which  compared the predicted velocities from the {\it IRAS } 1.2 Jy redshift surveys and the MARK III Tully-Fisher data, revealed systematic discrepancies that could not be attributed to 
errors in the data and the reconstruction methods.
Inspection of the flow fields obtained in the current work with the
those presented in DNW96  (see their figures 9--13) clearly show that the problem lies in the MARK III data set. The velocity fields predicted from the {\it IRAS} 
and {2MRS} surveys have similar patterns which grossly deviate from 
 MARK III  but are in accordance with SFI++. 
 The 2MRS  has all the attributes that one would want for estimating the gravity field including a very weak bias.  The survey was done by an instrument that was photometrically  stable,  which is important to avoid  large scale drifts in the derived gravity  field. The survey is far superior to the {\it IRAS} survey, the first full sky galaxy survey, which detected galaxies at $60\mu m$, a sign of star formation and not a good indicator of mass. However, that survey led to sensible results, and was not at fault for the disagreement 15 years ago.

\cite{radsm} compared the predicted velocities from the  {\it IRAS PSCZ} survey have also  been compared with measured peculiar velocities  of  SNe.  They found a best-fitting
   $\beta_{_{IRAS}}=f/b_{_{IRAS}}=  0.55 \pm 0.06$. The lower value of $\beta$ derived in our work could be 
   due to a difference in the biasing factor between 2MRS and the {\it IRAS} galaxies, but we also emphasize 
   that our estimation of the error in the predicted velocities  should be more reliable as it is based on realistic mock catalogs. 
   \cite{p05} performed a comparison of the 2MRS predicted velocities with 
direct velocity measurements from three different samples, including 836 SFI++ galaxies
within $cz=5000 \kms$. Their analysis yields $\beta=0.55 \pm .05$ for the comparison of gravity
with the SFI++. They derive $\Omega=0.55\pm 0.05$, inconsistent with our result at more than the 2.5$\sigma$ level.   However, they did not calibrate their methods with advanced 
mock catalogues nor included the expected covariance of the 
predicted velocities.  

\cite{l10} employed a sophisticated version of the nonlinear  MAK reconstruction method \citep{f02}
to compare the 2MRS  predicted velocities with the 3K velocity catalog
\citep{t08} of 1791 galaxies with 
redshifts $<3000\kms$.  They derive $\Omega=0.31\pm 0.05$, corresponding to $\beta \sim 0.52$. However, as they point out their error analysis is incomplete. Their method is promising as 
it takes intro account of nonlinear effects. Nevertheless,  they do not account for the covariance of the 
errors in their smoothed observed velocities and  predicted 
velocities.

Both \cite{p05} and \cite{l10} use iterative schemes based on \cite{y91} for
deriving the peculiar velocities from redshift surveys. 
These schemes rely on a relation between the peculiar velocity and density 
in real space.
 At any iteration, this relation is solved  for new peculiar velocity 
given  real space coordinates obtained from the observed  redshifts by subtracting  
 the old peculiar velocities derived in  the previous iteration. 
 We caution here that the these schemes are intrinsically biased:
error in velocities used to estimate the distances will 
yield a biased density field in real space (see the \S\ref{sec:note} for details). Hence the 
estimation of the velocity field is actually done from a biased distribution in real space. 
 The bias produces an undesired smoothing of  density field along the radial direction.
The smoothing width (in $\kms$)  is equal to the rms random error in the 
velocities $\sim 200-300 \kms$ \citep{be02, NussBranch}. Therefore,  the bias is more pronounced  in nonlinear methods  which aim at probing small scales.

  Checks to  find the best way of estimating the gravity field did not lead to improvements.  Weighting the galaxy maps by the 2MRS luminosity led to a worse agreement, and recall that 2MRS is selected in K band, which is closest to a measure of the stellar mass.  Giving the elliptical galaxies double weight, as indicated by lensing analysis \citep{mskh06}, did not improve the agreement.  This in itself is not too surprising, because on large scale the 2MRS survey is dominated by spiral galaxies.   It seems that the galaxies brighter than  $M_{*}+2$ 
 are each surrounded by a dark matter halo that has the same mass on average.  There is no hint that the dark matter mass is larger if the luminosity is increased.
  
  Using our estimate for  $\beta$ we can constrain  the amplitude of mass
fluctuations.  As a measure of the 
amplitude  we consider the rms of density fluctuations in spheres of $8\hmpc$ in radius, denoted
by $\sigma_8$ and $\sigma_{\rm 8g}$ for the mass and galaxy distributions, respectively. 
Adopting $\Omega=0.266$  \citep{wmap7} 
gives $f(\Omega,\Lambda=1-\Omega) = 0.483 $ \citep{lind05}. 
Comparing this to our result $\beta=f/b=0.33\pm 0.04 $ ($1\sigma$ error) we get   a bias factor  $b=1.46\pm .20$
between the dark matter and the 2MRS  galaxy distribution. 
Taking  $\sigma_{\rm 8g}=0.97\pm 0.05$  \citep{w09,rpe09},  yields  $\sigma_8=\sigma_{\rm 8g}/b=0.65 \pm 0.11$
  for the underlying mass density field,  marginally consistent  with the latest WMAP results
  \citep{wmap7} of $\sigma_8=0.8\pm 0.03$ (see also 
 \cite{jaro10}). 

\section{Conclusions}
We summarize the major conclusions of our work:

\begin{itemize}
\item{After a detailed examination of the 2MRS and SFI++ catalogs, we find the local gravity field  to be a fine predictor of the local velocity field. Such a conclusion is a comfort for linear perturbation theory in an expanding universe and was certainly expected. It is interesting that the counts of galaxies give the best possible gravity field, reinforcing the old idea that the mass of the halo around a galaxy is not very well correlated with the luminosity of that galaxy.}

\item{ We see no evidence that the dark matter does not follow the galaxy distribution,  and is consistent with constant bias on large scales. There is no evidence for a non-linear bias in the local flows. A smooth component to the universe is not something testable with these methods.}

\item{ Linear perturbation theory appears to be adequate for  the large scales  tested by our method.}
 
\item{The solution favors $\alpha=0.$, no correlation between luminosity and mass, and $\beta = .33 \pm .04$, which is consistent, but more than twice as tight as  \cite{e06a}. Using the derived $\Omega$ from WMAP \citep{jaro10}, leads to an estimate  $\sigma_8=0.65 \pm 0.10$, deviant from WMAP's reports at the $1.5\sigma$ level.}

\item{Our estimate of $\sigma_8 $ gives the most precise value at $z\sim 0$ and is useful for tests of the growth rate and Dark Energy.}

\item{The velocity-gravity comparison measures the acceleration on scales up to $30 - 50$ Mpc. and  since we derived a  similar value of $\beta$  as for clusters of galaxies, we conclude  that dark matter appears to fully participate in the clustering on scales of a few Megaparsecs and larger.} 
 
\item{We find no evidence for large-scale flows such as reported by,  for example  
\citep{hs99,he97,feldwh10}.  Note that our analysis has not used the CMBR dipole,  but we see a velocity field that is fully consistent with those previously reported \citep{e06, e06a, el09}, which are consistent with the  CMBR  dipole radiation. We see no evidence that the dipole in the CMBR is produced by anything other than our motion in the universe.}

\end{itemize}

\section{Acknowledgments}
We are all grieving over the untimely death of John Huchra, an old dear friend who first collaborated on papers describing the CfA1 survey. We are indebted to him for his contribution to this research, including being responsible for completion of the 2MRS and for pushing the observations to low galactic latitude. 

We thank Enzo Branchini for a careful reading of the manuscript.   MD acknowledges the support provided by the NSF grant  AST-0807630. 
The majority of this research was carried out at the MPA, Garching, when MD was supported by a Humboldt fellowship.   AN thanks the MPA for the hospitality. 
 This work was supported by THE ISRAEL SCIENCE FOUNDATION (grant No.203/09), the German-Israeli Foundation for 
Research and Development,  the Asher Space Research
Institute and  by the  WINNIPEG  RESEARCH FUND.
KLM acknowledges funding from the Peter and Patricia Gruber Foundation
as the 2008 Peter and Patricia Gruber Foundation International
Astronomical Union Fellow, from a 2010 Leverhulme Trust Early Career
Fellowship and from the University of Portsmouth and SEPnet
(www.sepnet.ac.uk).
The Millennium Simulation databases used in this paper and the web
application providing online access to them were constructed as part of
the activities of the German Astrophysical Virtual Observatory.

\bibliographystyle{mn2e}
\bibliography{refs_D}

\appendix

 \section{A note on iterative scheme}
 \label{sec:note}
We caution  of a possible systematic bias which may be important in 
iterative schemes for reconstructing velocities from the distribution of galaxies 
in redshift space. These schemes  rely on the  availability of a  relation between 
the peculiar velocity and density in real space, e.g., the linear relation 
$f {\rm div} {\bf v}=-\delta$. 
At the end of any iteration  intermediate peculiar velocities are provided, which are used 
as to derive the distances from the redshifts in the next iteration. Given those distances,  the adopted real space relation
between $\delta$ and $\bf v$ is then  solved to obtain a new guess for the velocities. 
The loop 
is continued until the change in the peculiar velocity between successive iterations becomes
smaller than a certain threshold. Vanishing peculiar velocities could be taken as input for the first iteration. 

We demonstrate here that the real space distribution of galaxies as obtained from the output from any iteration scheme is 
biased. Hence the corresponding peculiar velocity is also biased. 
We will first show that a biased distribution in real space is obtained even if 
unbiased but noisy peculiar velocities are used to get the distances. 
We write the density of galaxies in real space at distance $r$ in a given direction on the sky  as
\begin{equation}
n_{\rm e}(r) r^2=(2\pi \sigma^2)^{-1/2}\int \dd s s^2 n_{\rm s} {\rm e}^{-\frac{[s-v(s)-r]^2}{2\sigma^2}} \; ,
\end{equation}
where $s=r+v$  is the radial redshift space coordinate, $v(s)$ is the peculiar velocity of a
galaxy present at $ s$, and $\sigma$ is the rms of the error in the determination of the $v(s)$. In the above we assume normal error distribution and  a one-to-one mapping between $s$ and $v$, i.e. 
we neglect fingers-of-god effects and triple value zones. Working with the variable $r_1=s-v(s)$ we get
\begin{equation}
n_{\rm e}(r) r^2=(2\pi \sigma^2)^{-1/2}\int \dd r_1 r_1^2 n_{\rm t}(r_1)  {\rm e}^{-\frac{(r_1-r)^2}{2\sigma^2}} \; ,
\label{eq:r1}
\end{equation} 
where 
\begin{equation}
n_t(r_1,)= \frac{\dd s}{\dd r_1} \left[1+\frac{v(s(r_1))}{r_1}\right]^2 n_{\rm s}(s(r_1)) \; , 
\label{eq:a2}
\end{equation}
is the actual real space density at $r_1$.
Therefore, errors in the peculiar velocities (even if unbiased  relative to the true ones) 
cause  a smearing of structure in the radial direction. 
This anisotropic smearing is important for 
scales   $\sim  \sigma$ (in $\kms$).  
The bias is similar to the traditional inhomogeneous Malmquist bias which is usually encountered in studies of  distance indicators \citep{lyn88}. 
 
A self-consistent treatment of the bias should take into account the fact that 
$v(s)$ used in  \ref{eq:r1}
 and \ref{eq:a2} is the biased peculiar velocity obtained from $n_{\rm e}$.
This could be done in the far observer limit, $|v/r|<<1$, and for small perturbations
where equation  \ref{eq:a2} reduces to 
\begin{equation}
\delta_{\rm t}=\delta_{\rm s}+\frac{\dd v}{\dd s}\; .
\end{equation}
Substituting this into \ref{eq:r1} and   Fourier transforming the result we get
\begin{equation}
\tilde \delta_{\rm e}({\bf k})=\left[\tilde \delta_{\rm s} ({\bf k}) + i k_r  \tilde v({\bf k}) \right ]
{\rm e}^{-k_r^2 \sigma^2/2}\; ,
\end{equation} 
where $k_r$ is the component of of $\bf k$ parallel to the line of sight and the tilde denotes 
quantities in k-space.
Using this last equation in  the linear $\delta - v$ relation,  $ f \tilde v=-i (k_r/k^2) \tilde \delta_{\rm e} $, 
we find
\begin{equation}
\tilde \delta_{\rm e}({\bf k}) =\frac{\tilde \delta_{\rm s} ({\bf k}){\rm e}^{-k_r^2 \sigma^2/2} }{1+f (k_r/k)^2{\rm e}^{-k_r^2 \sigma^2/2}}\; ,
\end{equation}
instead of the usual unbiased expression obtained with $\sigma=0$.

A more complete analysis of the bias must incorporate the covariance of the errors in the 
derived peculiar velocity field.

\end{document}